\documentclass[final]{article}

\usepackage[T1]{fontenc}
\usepackage{fourier}
\usepackage[letterpaper,margin=1in]{geometry}

\usepackage{mathtools}
\usepackage{amssymb}
\usepackage{bm}
\usepackage[mathscr]{eucal}

\usepackage{array}
\usepackage{booktabs}
\usepackage{tabularx}
\usepackage{caption}
\usepackage{color}
\usepackage[inline]{enumitem}
\usepackage{fancyhdr}
\usepackage{graphicx}
\usepackage{setspace}
\usepackage[subrefformat=parens]{subcaption}
\usepackage{threeparttable}
\usepackage{titlesec}
\usepackage{url}
\usepackage{xcolor}
\usepackage[multiple]{footmisc}

\usepackage[most]{tcolorbox}
\makeatletter
\newtcolorbox{nicebox}[1][]{
  colback=white, colframe=black, sharp corners,after app={\@endparenv},#1}
\makeatother

\PassOptionsToPackage{natbib,maxcitenames=2}{biblatex-chicago}
\usepackage[authordate, backend=biber, uniquename=false, noibid]{biblatex-chicago}

\DeclareFieldFormat{citehyperref}{%
  \DeclareFieldAlias{bibhyperref}{noformat}
  \bibhyperref{#1}}

\DeclareFieldFormat{textcitehyperref}{%
  \DeclareFieldAlias{bibhyperref}{noformat}
  \bibhyperref{%
    #1%
    \ifbool{cbx:parens}
      {\bibcloseparen\global\boolfalse{cbx:parens}}
      {}}}

\savebibmacro{cite}
\savebibmacro{textcite}

\renewbibmacro*{cite}{%
  \printtext[citehyperref]{%
    \restorebibmacro{cite}%
    \usebibmacro{cite}}}

\renewbibmacro*{textcite}{%
  \ifboolexpr{
    ( not test {\iffieldundef{prenote}} and
      test {\ifnumequal{\value{citecount}}{1}} )
    or
    ( not test {\iffieldundef{postnote}} and
      test {\ifnumequal{\value{citecount}}{\value{citetotal}}} )
  }
    {\DeclareFieldAlias{textcitehyperref}{noformat}}
    {}%
  \printtext[textcitehyperref]{%
    \restorebibmacro{textcite}%
    \usebibmacro{textcite}}}

\AtBeginBibliography{

}

\usepackage[hidelinks,
  colorlinks=true,
  citecolor=black,
  linkcolor=darkblue,
  urlcolor=magenta]{hyperref}
\usepackage{doi}

\usepackage{zref-clever}
\zcsetup{nameinlink=false}
\zcLanguageSetup{english}{cap,rangesep=\textendash}
\zcRefTypeSetup{equation}{nocap,Name-sg=Eq.,name-sg=,Name-pl=Eqs.,name-pl=,}
\NewDocumentCommand{\cref}{m}{\zcref{#1}}

\NewDocumentCommand{\Cref}{m}{\zcref[S]{#1}}
\NewDocumentCommand{\crefrange}{mm}{\zcref[range,rangetopair=false]{#1,#2}}
\NewDocumentCommand{\Crefrange}{mm}{\zcref[S,range,rangetopair=false]{#1,#2}}

\usepackage{graphicx}
\usepackage{overpic}
\usepackage{tikz}
\usetikzlibrary{positioning}
\usetikzlibrary{shapes.geometric}
\usetikzlibrary{shapes}
\usetikzlibrary{trees}

\usepackage{keytheorems}
\newkeytheorem{proposition}[style=plain,qed=\textup{\guillemotleft}]

\definecolor{darkblue}{rgb}{0,0,0.55}
\definecolor{darkred}{rgb}{0.5,0,0}
\newcommand\Bheadfont{\fontsize{14pt}{\baselineskip}\selectfont}

\definecolor{darkblue}{rgb}{0,0,0.55}
\definecolor{darkred}{rgb}{0.5,0,0}

\titleformat{\section}[hang] {\normalfont\sc\color{darkblue}\Bheadfont} {\thesection\hskip0.618em}{0em}{}
\titlespacing*{\section}
{0pt}{15pt plus 2pt minus 2pt}{9pt plus 2pt minus 2pt}
\titleformat{\subsection}[runin]
{\normalfont\sc\color{darkblue}} {\thesubsection\hskip0.618em}{0em}{}
\titlespacing*{\subsection}
{0pt}{13pt plus 2pt minus 2pt}{13pt plus 2pt minus 2pt}
\titleformat{\subsubsection}[runin]
{\normalfont\sc\color{darkblue}} {\thesubsubsection\hskip0.618em}{0em}{}
\titlespacing*{\subsubsection}
{0pt}{13pt plus 2pt minus 2pt}{13pt plus 2pt minus 2pt}

\DeclarePairedDelimiter{\set}{\lbrace}{\rbrace}
\DeclarePairedDelimiter{\abs}{\lvert}{\rvert}

\newcommand{\E}{\mathbb{E}}
\newcommand{\diff}{\mathrm{d}}
\newcommand{\defas}{\coloneq}

\newcommand{\bt}{\mathbf{t}}
\newcommand{\br}{\mathbf{r}}

\NewDocumentCommand{\Expect}{om}{\E\IfValueT{#1}{_{#1}}\IfValueF{#1}{\!}\left[#2\right]}
\NewDocumentCommand{\Var}{om}{\mathrm{Var}\IfValueT{#1}{_{#1}}\IfValueF{#1}{\!}\left[#2\right]}
\NewDocumentCommand{\Prob}{om}{\Pbb\IfValueT{#1}{_{#1}}\IfValueF{#1}{\!}\IfBlankF{#2}{\left(#2\right)}}

\newcommand{\gpkern}[1]{\textrm{#1}}
\newcommand{\gpC}{\gpkern{Constant}}
\newcommand{\gpLin}{\gpkern{Linear}}
\newcommand{\gpSe}{\gpkern{SquaredExponential}}
\newcommand{\gpPer}{\gpkern{Periodic}}
\newcommand{\gpGe}{\gpkern{GammaExponential}}

\newcommand{\pC}{\gpkern{C}}
\newcommand{\pLin}{\gpkern{LIN}}
\newcommand{\pSe}{\gpkern{SE}}
\newcommand{\pGe}{\gpkern{GE}}
\newcommand{\pPer}{\gpkern{P}}

\newcommand{\gpSum}{+}
\newcommand{\gpMul}{\times}

\newcommand{\pSum}{+}
\newcommand{\pMul}{\times}

\newcommand{\SGP}{Seasonal AutoGP}

\makeatletter
\newcommand{\zcBookmarkNumber}[1]{%
  \zref@extractdefault{#1}{default}{??}%
}
\newcommand{\zcBookmarkTypeName}[1]{%
  \@ifundefined{zcBookmarkType@\zref@extractdefault{#1}{zc@type}{}}%
    {\zref@extractdefault{#1}{zc@type}{??}}%
    {\@nameuse{zcBookmarkType@\zref@extractdefault{#1}{zc@type}{}}}%
}
\newcommand{\zcBookmarkType@proposition}{Proposition}
\newcommand{\titlezcref}[1]{%
  \texorpdfstring
    {\zcref[S]{#1}}%
    {\zcBookmarkTypeName{#1}~\zcBookmarkNumber{#1}}%
}
\makeatother

\begin{document}

\title{\Large{Probabilistic Seasonality}}
\author{
Feras A.~Saad%
  \thanks{Computer Science Department, Carnegie Mellon University, \href{mailto:fsaad@cmu.edu}{fsaad@cmu.edu}.}
\and
Todd B.~Walker%
  \thanks{Department of Economics, Indiana University, \href{mailto:walkertb@iu.edu}{walkertb@iu.edu}}}
\date{September 2026}
\maketitle

\vspace{0.7in}

\begin{abstract}
Seasonal adjustment is fundamental to economic analysis,
but uncertain because seasonal components are inherently latent.
This article introduces a probabilistic model discovery method that
decomposes a time series into seasonal and nonseasonal components.
The method returns a posterior distribution over the structure and parameters
of a seasonal component.
In simulation studies, the method can improve point forecasts, interval
predictions, and recovery of seasonal components relative to
X-13ARIMA-SEATS.
In a study of eight U.S.~macroeconomic series during the COVID-19
recession, the method surfaces significant ex-ante uncertainty about
current seasonal adjustments in real time, well before many
X-13 revisions reach their eventual peaks.
\end{abstract}

\setcounter{page}{0}
\thispagestyle{empty}

\clearpage
\onehalfspacing

\section{Introduction}
\label{Sec:Introduction}

Many economic time series, such as retail sales, unemployment, and
inflation, exhibit strong seasonal patterns that are
driven by calendar effects, weather, holidays, and institutional factors.
The magnitude of this within-year variation can even exceed the impact of
recessions.
In nonfarm payroll employment, for example, the average absolute difference
between seasonally adjusted and unadjusted monthly changes is much
larger than the typical month-to-month variation in the adjusted data
\citep{Wright2013}.
The removal of recurring fluctuations allows analysts, policymakers, and
forecasters to distinguish underlying trends, business-cycle movements, and
irregular shocks more clearly.
Accurate seasonal adjustment is therefore essential for real-time
monitoring of economic conditions, constructing reliable leading
indicators, and evaluating policy.
These issues have received renewed attention following unusually large downward
revisions, released in August 2025, to U.S.~nonfarm payroll employment
changes for May and June by a combined 258,000 jobs~\citep{BLS2025EmploymentJuly}, which
prompted heightened public scrutiny of the production and revision of
official labor-market statistics~\citep{Mutikani2025}.

Distinguishing seasonal variation from cyclical and irregular movements has
nevertheless remained a fundamental problem, as formally articulated in
\citet{Nerlove:1964}.
Several recent observations make the point.
\citet{RudebuschEtAl:2015} document systematically weak first-quarter GDP
growth that was subsequently attributed to residual seasonality surviving
in the published seasonally adjusted aggregates
\citep{Lunsford:2017,ConsolvoLunsford:2019}.
\citet{Lunsford:2025} finds a similar pattern in five measures of PCE
inflation.
At the onset of the COVID-19 pandemic, the collapse in economic activity
coincided with the seasonal trough in the winter housing market, leading
standard procedures to overstate the level of housing starts immediately
before the collapse and therefore the apparent depth of the subsequent
contraction \citep{BrysonCornwall:2026}.
These examples illustrate that seasonal components are inherently
latent and can exhibit significant uncertainty.

Conventional approaches to seasonal adjustment
typically report only a conditional mean, with little-to-no
information about revisions, turning points, and cyclical contamination.
Major seasonal adjustment procedures used by government agencies---such as
X-13ARIMA-SEATS---perform a battery of frequentist specification tests
to identify and remove seasonality \citep{X13RefManual:2024}.
While each test may appear statistically valid in isolation, the overall
procedure lacks a unified inferential framework.
Moreover, the procedure delivers only point estimates of seasonal factors.
Once these estimates are obtained, uncertainty about the decomposition itself is
typically discarded or treated as a secondary diagnostic.

We introduce an alternative approach based on a probabilistic view of
seasonality.
Our framework rests on Bayesian model discovery within a family of
Gaussian process models~\citep{Rasmussen:2006}
that can express a variety of time series patterns.
The method does not impose a particular form of seasonality a-priori, but
instead infers the structure and parameters of periodic patterns---such as
additive, multiplicative, or locally varying periodicity---from observed data.
The use of periodic structures is consistent with the usual
characterization of seasonality as a component whose pattern repeats from
year to year~\citep{Granger:1978}.
We term the resulting method, which is based on the Automatic Gaussian
Process (AutoGP) method of \citet{SaadEtAl:ICML:23},
\textit{\SGP}.

Given a time series dataset,
\SGP{} infers a posterior distribution over its decomposition
into seasonal and nonseasonal components.
This probabilistic representation supports real-time updates to the model
as data becomes available.
It also enables direct posterior queries without auxiliary specification
tests or statistical re-estimation of the entire model on a per-query
basis.
In particular, the method
\begin{enumerate*}[label=(\roman*)]
\item discovers the presence and form of periodicity rather than imposing it;
\item returns a full posterior over seasonal paths rather than a point factor; and
\item quantifies existence, strength, sign and timing, model disagreement, and revision risk
associated with seasonal components.
\end{enumerate*}
The framework reports seasonally adjusted values along with
an ex-ante measure of how much confidence the analyst should place in
the decomposition, before its eventual revision is observed.

We first assess the efficacy of the framework in a simulation study that
mimics classical economic benchmarks, where the true seasonal component
is known.
The data-generating process is calibrated to the \citeauthor{BoxJenkins:TimeSeries:1976}
data of monthly international air passengers from 1949--1960.
This dataset remains the
canonical benchmark for seasonal adjustment and supplies the default
$(0,1,1)(0,1,1)_{12}$ ``airline model'' specification of X-13ARIMA-SEATS
itself \citep{BoxJenkins:TimeSeries:1976,X13RefManual:2024},
so the simulation is intentionally staged on terrain that favors X-13.
Fitting a sinusoid with linearly growing amplitude to the airline data
captures the upward trend, annual seasonal pattern, and widening
peak-to-trough variation of the series.
For the observed series, \SGP{} produces lower prediction errors
than X-13 across nearly all forecast horizons and sample
lengths, and its prediction intervals exhibit improved coverage
and sharpness across most experimental conditions.
The experimental results on predicting the seasonal component reveal an
informative tradeoff between imposing seasonality (as in X-13) and
discovering it from the data (as in \SGP).
When the seasonal signal is weak, neither method dominates.
At intermediate signal-to-noise ratios, the imposed seasonal structure
from X-13 is more accurate.
In the longest samples, where the data contain sufficient information to
identify periodic structure, \SGP{} reduces prediction
error at ten of the twelve forecast horizons.

We then study eight monthly U.S. macroeconomic series spanning labor
markets, consumer prices, industrial production, retail spending,
residential construction, manufacturing demand, and business inventories.
Compared to X-13,
\SGP{} produces lower twelve-month not-seasonally-adjusted
point forecast errors for six of the eight series,
and a lower interval forecast error in seven.
The largest reductions include 33 percent for core
CPI and 24 percent for retail sales.
Both \SGP{} and X-13 nonetheless remove similar amounts of conventional
fixed-frequency seasonal variation.
In most cases, the residual seasonal-harmonic power is at most a few
percent of the unadjusted power under either procedure.
These results show that \SGP{}'s forecast improvements do not occur at the
cost of materially weaker removal of conventional fixed-frequency
seasonality.

Our final application compares the extent to which X-13 and \SGP{}
respond to large changes that occurred in these eight time series
during the COVID-19 recession.
We find that X-13 and \SGP{} produce comparable revisions, but
the former produces revisions much later in the sample period.
Estimates from X-13 show that the seasonal components initially assigned to
the COVID-19 shock month were subsequently revised by
multiples of their ordinary historical revisions, and typically only after
long delays.
For example, peak revisions from X-13 arrive 37 months later for payroll
employment and housing starts, 33 months later for retail sales, and at
least ten months later for every series except durable-goods orders.
Because these revisions become observable only as later data arrive,
conventional revision diagnostics identify the instability ex post.

Using \SGP{}, we exploit the probabilistic nature of seasonal adjustments
to define a so-called ``Revision Risk'' distribution, which provides a
coherent measure of the uncertainty in a seasonal adjustment at the time of
release.
Informally, the Revision Risk answers the question: \textit{How large are
future revisions to the current seasonal adjustment likely to be?}
The Revision Risk from \SGP{} already exceeds its historical 95th
percentile for four of eight series at the COVID-19 shock month itself, and
for all eight just three months later, when the median index is 8.1 times
the historical median.
While X-13 and \SGP{} eventually agree about the seasonal pattern---with
a median correlation between the seasonal components of 0.972---the timing
and uncertainty of the adjustments are distinct.
\SGP{} makes most of the adjustment between two and five months
after the shock and reports the associated uncertainty in real time.
In contrast, X-13 defers most rewriting for one to three years later, and
offers no analogous ex-ante probabilistic distribution of future revisions
to seasonal adjustments.
This study underscores that a central contribution of the probabilistic
approach is its ability to monitor the uncertainty in adjustments in real
time, instead of treating uncertainty using retrospective diagnostics as in X-13.

The remainder of this paper is structured as follows: \cref{sec:discovery}
describes probabilistic model discovery and AutoGP; \cref{sec:seasonality}
introduces its extension to \SGP{}; \cref{sec:simulation} presents a
simulation study on airline data; \cref{sec:empirical} analyzes eight
U.S.~economic datasets; and \cref{sec:conclusion} concludes the paper.

\section{Probabilistic Model Discovery}
\label{sec:discovery}

We begin with a description of our approach to learning probabilistic
models of time series data.

\subsection{Bayesian Formulation}
\label{sec:discovery-formulation}

Let $\mathcal{M} = \set{m_1, m_2, \ldots }$ denote a set of
statistical models.
Each model $m \in \mathcal{M}$ defines a likelihood function
$f(\mathbf{y};  \theta, m)$ for observable data
$\mathbf{y}$, parameterized by $\theta \in \Theta_{m}$.
Using a prior distribution $\pi(m)$ over models  along with a prior
distribution $\pi(\theta | m)$ over parameters within a model induces the
joint distribution
\begin{equation}
p(m, \theta, \mathbf{y}) = \pi(m) \pi(\theta | m) f(\mathbf{y}; \theta, m),
\label{eq:joint-dist-mty}
\end{equation}
and in turn the following posterior distributions:
\begin{enumerate}[wide,label={\roman*.}]
\newcommand{\eqcond}[1]{\mathmakebox[30em][l]{#1}}

\item over models and parameters,
\begin{equation}
\label{eqn:PosteriorModelsParams}
\mathmakebox[.6\displaywidth][l]{%
  p(m,\theta \mid \mathbf{y})
  =
  \frac{\pi(m)\pi(\theta \mid m)f(\mathbf{y};\theta,m)}
  {\sum_{m' \in \mathcal{M}} \int_{\Theta_{m'}}
    \pi(m')\pi(\theta' \mid m')f(\mathbf{y};\theta',m')
    \,\diff\theta'}}
\mathrlap{(m \in \mathcal{M},\ \theta \in \Theta_m)};
\end{equation}

\item over the model space,
\begin{equation}
\label{eq:PosteriorModelSpace}
\mathmakebox[.6\displaywidth][l]{%
  p(m \mid \mathbf{y})
  =
\int_{\Theta_m} p(m,\theta \mid \mathbf{y}) \-,\diff\theta}
\mathrlap{(m \in \mathcal{M})};
\end{equation}

\item the posterior over parameters, for specific model $m^\star \in \mathcal{M}$,
\begin{equation}
\label{eq:PosteriorParams}
\mathmakebox[.6\displaywidth][l]{%
  p(\theta \mid \mathbf{y}, m^\star)
  =
  \frac{\pi(\theta \mid m^\star)f(\mathbf{y};\theta,m^\star)}
  {\int_{\Theta_{m^\star}}
    \pi(\theta' \mid m^\star)f(\mathbf{y};\theta',m^\star)
    \,\diff\theta'}}
\mathrlap{(\theta \in \Theta_{m^\star}).}
\end{equation}

\end{enumerate}
As the distributions of interest in \crefrange{eqn:PosteriorModelsParams}{eq:PosteriorParams}
are almost always intractable to compute exactly, they are instead
approximated using statistical algorithms
such as Markov chain Monte Carlo or sequential Monte Carlo~\citep{Robert:2004}.
These algorithms
generate a set $\mathcal{S} = \set*{ \left(w^{(i)}, \left(m^{(i)}, \theta^{(i)}\right)\right) }_{i=1}^M$
of $M$ weighted samples that form a discrete approximation of the target distribution.
These samples should be properly weighted~\citep[\S2.5.4]{Liu:2004} for the posterior,
meaning that for any square integrable function $\varphi$ over models, parameters
and data,
\begin{align}
\Expect{w^{(i)}\varphi\left(m^{(i)},\theta^{(i)}, \mathbf{y}\right) \;\middle|\; \mathbf{y}}
  = c \; \Expect[p(m,\theta \mid \mathbf{y})]{\varphi(m, \theta, \mathbf{y})}
  && (1 \le i \le M),
\end{align}
where $c > 0$ is a constant common to all $M$ samples.
Equipped with properly weighted samples,
consistent estimates of the expectation of $\varphi$
can be computed via ratio estimation
\begin{equation}
  \Expect[p(m,\theta \mid \mathbf{y})]{\varphi(m, \theta, \mathbf{y})}
  \approx \sum_{i=1}^M \left( \frac{w^{(i)}}{\sum_{j=1}^M w^{(j)}} \right) \varphi\left(m^{(i)}, \theta^{(i)}, \mathbf{y}\right).
\end{equation}
This framework enables flexible inference over both model structures and
parameters and forms the basis for structural and numeric queries.
A key challenge is designing a model space $\mathcal{M}$ that is both
expressive and interpretable.
Following \cite{SaadEtAl:ICML:23}, our approach constructs $\mathcal{M}$ as
a probabilistic context-free grammar (PCFG) over \textit{Gaussian process}
(GP) kernels, which describe the covariance structure of latent functions.

\subsection{Gaussian Processes}
\label{sec:discovery-gp}
A Gaussian process (GP) over an index set $\mathbb{T}$ is a family
$\set{ X(t) : t \in \mathbb{T}}$ of random variables
such that for any
time points
$\bt \defas (t_1 ,\dots,t_n) \in \mathbb{T}^n$,
the random vector
$X(\bt) \defas \left(X(t_1), \dots, X(t_n)\right)$
follows a multivariate Gaussian distribution:
\begin{equation}
  X(\bt) = \left(X(t_1), \dots, X(t_n)\right)
  \sim \mathcal{N} \left( \mu(\bt), K(\bt,\bt) \right).
\end{equation}
The symbol $\mu(\bt) \defas (\mu(t_1), \ldots, \mu(t_n))$
denotes the mean vector with
$\mu(t) \defas \Expect{X(t)}$, and $K(\bt,\bt)$
is the $n \times n$ covariance matrix whose entries are defined by a
\textit{covariance kernel}:
$[K(\bt,\bt)]_{ij} = k(t_i, t_j) \defas \Expect{(X(t_i) - \mu(t_i))(X(t_j) - \mu(t_j))}$
for all $1 \le i,j \le n$.
Conditioned on $X(\bt) = x(\bt)$,
the posterior distribution of
$X(\bt')$ at new time points
$\bt' = (t'_1, \ldots, t'_{n'}) \in \mathbb{T}^{n'}$,
is also a multivariate Gaussian:
\begin{equation}
X(\bt') \;\big\vert\; X(\bt) = x(\bt) \sim \mathcal{N}\left(\mu^{\text{post}}(\bt'), K^{\text{post}}(\bt', \bt')\right),
\end{equation}
which forms the basis of predictive inference on unobserved data given
observed data $X(\bt)$.
The specific forms of $\mu^{\rm post}$ and $K^{\rm post}$,
as well as further background on Gaussian processes,
is given in \cref{appx:gp-model}.

\subsection{Covariance Kernels}
\label{sec:discovery-kernels}

\begin{figure}[!h]
\includegraphics[width=\linewidth]{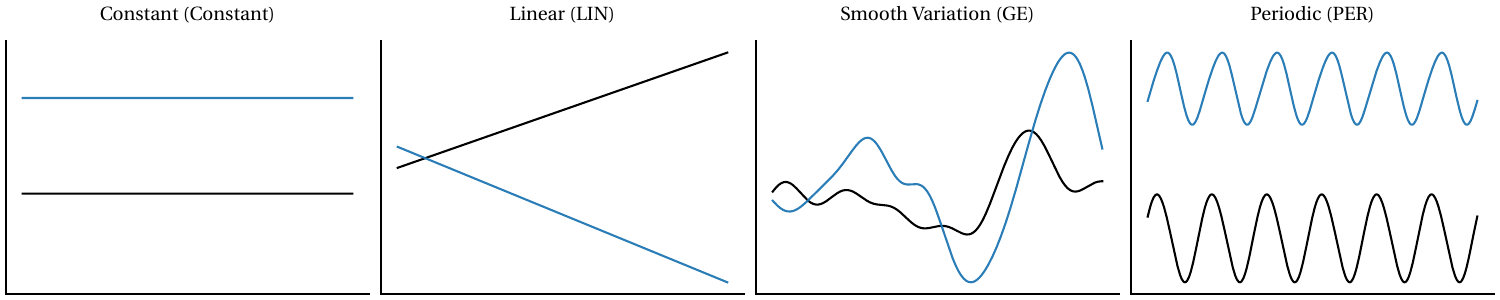}
\caption{Samples from Gaussian process models with a single covariance kernel.}
\label{fig:gp-single}
\end{figure}

Gaussian processes can be used to model time series models that exhibit
rich temporal characteristics, by appropriate selection of the \textit{covariance kernel}
$k: \mathbb{T} \times \mathbb{T} \to \mathbb{R}$ that defines the covariance
matrix $K$ of the underlying multivariate normal.
Covariance kernels provide a compact representation of a potentially
infinite-dimensional feature space---rather than explicitly specifying basis
functions, the covariance kernel $k$ encodes the
similarity between inputs $(t, t')$,
inducing a rich space of functions \citep{Rasmussen:2006}.
We will consider the following kernels, whose full definitions are given
in \cref{appx:gp-kernels}:
$\textbf{\gpC}$ for modeling constant functions,
$\textbf{\gpLin}$ for modeling linear functions,
$\textbf{\gpGe}$ for modeling smoothly varying functions,
and
$\textbf{\gpPer}$ for modeling periodically varying functions.
\Cref{fig:gp-single} shows representative samples from Gaussian processes
with these kernels, excluding measurement noise.

\subsection{Kernel Composition}
\label{sec:discovery-composition}

More expressive Gaussian process models can be constructed by composing
primitive covariance kernels to form composite kernels.
Given a sufficiently expressive set of primitive kernels and composition operators,
Gaussian processes can serve as universal approximators for continuous
functions on compact sets~\citep{Micchelli:2006}.
Two covariance kernels $k_1$ and $k_2$ can be combined
to create a new kernel through the following operations:
\begin{itemize}[noitemsep]
  \item \textbf{Addition} $k_\pSum(t,t') \defas k_1(t,t') + k_2(t,t')$.
  This operator supports additive structure, such as a seasonal component superimposed on a trend.

  \item \textbf{Multiplication} $k_\pMul(t,t') \defas k_1(t,t') \pMul k_2(t,t')$.
  This operator allows for modulation, such as a periodic component with
  time-varying amplitude.
\end{itemize}

\begin{figure}[t]
\includegraphics[width=\linewidth]{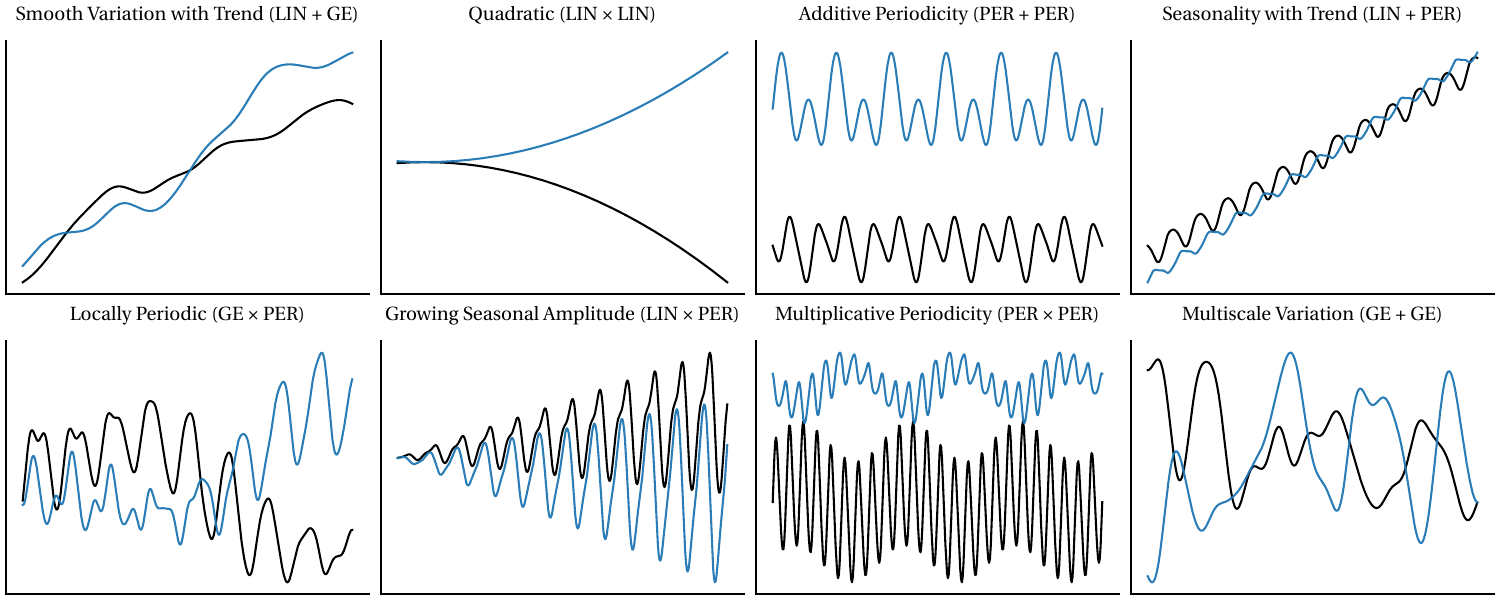}
\caption{Samples from Gaussian process models with composite covariance kernels.
The models can express a variety of time series structures, by varying the
primitive covariance kernels and combining them using addition and multiplication.}
\label{fig:gp}
\end{figure}

\Cref{fig:gp} shows examples of Gaussian process time series data
that arise from using different composite covariance kernels,
demonstrating a range of structural patterns that the model can express.

\subsection{Automated Kernel Discovery}
\label{sec:discovery-search}

Recent advances have made it possible to discover the structure of a
time-series dataset, by automatically learning the covariance kernel of a
Gaussian process model for the data
\citep{Duvenaud:ICML:2013,Saad:Dissertation:22,SaadEtAl:ICML:23}.
Rather than requiring the econometrician to manually specify a kernel,
these methods define a space of composite kernels via a symbolic grammar
that defines a countable model class $\mathcal{M} = \set{m_1, m_2, \dots}$.
This approach allows the econometrician to place a prior distribution over
model structures, not just parameters, enabling Bayesian structure learning
to perform inference over a space of Gaussian process models.
By doing so, it provides both expressivity and interpretability as each
sampled model can be inspected as a symbolic expression revealing
qualitative features of the time series, such as
seasonal variation, trends, and their interactions.
Following \citet{SaadEtAl:ICML:23}, we define a
prior over model structures $m \in \mathcal{M}$
using a \emph{probabilistic context-free grammar}
(PCFG), whose terminal symbols are primitive covariance kernels and whose
production rules leverage kernel composition operators:
\begin{align}
B                                            & \sim \gpC \mid \gpLin \mid \gpGe \mid \gpPer                && (\mbox{w.p. } \alpha_1, \alpha_2, \alpha_3, \alpha_4) \label{eq:hbm-base}\\
\oplus                                       & \sim \gpSum \mid \gpMul                                     && (\mbox{w.p. } \beta, 1-\beta)              \label{eq:hbm-op}\\
m                                            & \sim B \mid \oplus[m_1,m_2]                                 && (\mbox{w.p. } \delta, 1 - \delta).                     \label{eq:hbm-m}
\end{align}
In the rule for $m$, the probability $\delta>1/2$ of selecting a primitive kernel
is chosen to ensure that the prior halts almost surely~\citep[\S4.2]{SaadEtAl:POPL:19}.
Equipped with the structure, we next define a prior over numerical
parameters of the covariance kernel, which fully defines the prior Gaussian
process:
\begin{align}
\theta_i \mid m &\sim p_{m,i} && (1 \le i \le d(m)) \label{eq:hbm-theta} \\
X(\cdot) \mid m,\theta              & \sim \mathrm{GP}\left(0, k_{m,\theta}\right) \label{eq:hbm-x}.
\end{align}
In \cref{eq:hbm-theta}, $d(m)$ denotes the number of numerical parameters
within the kernel structure $m$, e.g., $d\left(\gpPer\right) = 3$
and $d\left(\gpPer \gpSum \gpC \right) = 4$, and $p_{m,i}$ denotes a
prior over the parameters of the $i$th scalar parameter within $m$
(lognormal for all parameters, except for the $\gamma \in (0,2]$ parameter
of the $\gpGe$ kernel which uses a logit normal).
In \cref{eq:hbm-x}, $X(\cdot)$ is the (infinite-dimensional) latent
Gaussian process and $k_{m,\theta}$ denotes its covariance function, which
is determined by the sampled structure $m$ and parameters $\theta$.

The final step is to define an observation model.
Economic time series are often subject to measurement noise.
We model noisy data by using a Gaussian process
$Y(t) \defas X(t) + \epsilon(t)$
that is the sum of a latent Gaussian process and i.i.d.~Gaussian
noise, where $\epsilon(t) \sim \mathcal{N}(0, \eta)$ for $t \in \mathbb{T}$.
Assuming an inverse gamma prior over the noise variance $\eta$,
for any time points $\bt = (t_1, \ldots, t_n)$ the observable
noisy data $Y(\bt) \defas (Y(t_1), \dots, Y(t_n))$ is defined by
\begin{align}
\eta                          & \sim \text{InvGamma}(1,1) \label{eq:hbm-noise}\\
Y(t_i) \mid X(\cdot)          & \sim \mathcal{N}(X(t_i), \eta) && (i = 1, \ldots, n).                                   \label{eq:hbm-y}
\end{align}
We have thus defined all the ingredients for the joint distribution
\cref{eq:joint-dist-mty} over structures, parameters, and data.

\subsection{Posterior Inference}
\label{sec:discovery-inference}

Recall that our goal is to produce
samples $\set*{\left(w^{(i)}, \left(m^{(i)},\theta^{(i)}, \eta^{(i)}\right)\right)}_{i=1}^M$,
called a particle collection, of the hidden variables given an observation $Y(\bt) = y(\bt)$.
As the model does not exhibit conjugacy that enables closed-form posterior
inference, more sophisticated sampling methods are required to
explore the posterior $p(m, \theta, \eta \mid y(\bt))$.
We use AutoGP\footnote{\url{https://probsys.github.io/AutoGP.jl}},
which is implemented in the Gen probabilistic programming system~\citep{CusumanoTownerEtAl:PLDI:19}
to conduct approximate posterior inference.
The implementation uses sequential Monte Carlo
\citep{ChopinEtAl:2020} and involutive Markov Chain Monte Carlo
\citep{NeklyudovEtAl:ICML:2020} algorithms;
see \citet[\S3]{SaadEtAl:ICML:23} for full technical details.
Our contribution extends this inference engine to enable
probabilistic identification, decomposition, and adjustment of seasonal
data built on top of the posterior draws, as described in \cref{sec:seasonality} .
For brevity, however, the figures and discussion below label the resulting
estimates ``AutoGP,'' because this engine generates the particles
underlying every estimate we report.

\section{Probabilistic Seasonality}
\label{sec:seasonality}

We now extend the AutoGP method from \cref{sec:discovery} to
\SGP{}, which identifies seasonality from the learned Gaussian
process models and supports posterior queries.
Periodic structure is a defining characteristic of seasonality
in time series data~\citep{Granger:1978}.
Our framework aligns with this notion by defining seasonality in terms of
\gpPer{} kernels in the posterior over Gaussian process models.
By grounding the notion of seasonality in the kernel algebra, we obtain an
operational and model-based criterion that is amenable to
Bayesian inference.
We first give some brief intuition about the periodic kernel, following
\citet[\S5.2]{MacKay1998}.

\subsection{Periodic Kernel}
\label{sec:seasonality-kernel}

The periodic kernel imposes that the covariance between
observations should depend only on their \emph{phase difference} within a
fixed period $\rho$, rather than the absolute time index.
For example, monthly seasonality would imply strong correlation between, say,
January 2020 and January 2023, while January and July should be minimally correlated.
The \gpPer{} kernel, $k_\pPer$, is designed to capture such structure.
It can be derived from the \gpSe{} kernel,
which measures the similarity between two time points $(t, t')$
using the Euclidean distance $\Delta_{\text{eu}}(t,t') = \abs{t - t'}$,
\begin{equation}
  k_{\pSe}(t, t')
    = \sigma^2 \exp\left( -\frac{1}{2\ell^2}{\left(\Delta_{\text{eu}}(t,t')\right)^2} \right)
    = \sigma^2 \exp\left( -\frac{1}{2\ell^2}{\abs{t - t'}^2} \right),
    \label{eq:cov-se}
\end{equation}
where $\ell$ is the lengthscale parameter controlling smoothness and
$\sigma^2$ is the marginal variance.
While this kernel is stationary and smooth, it does not encode periodicity,
because correlations decay monotonically with absolute distance.
To enforce periodicity with period $\rho$, the time domain is wrapped onto the unit circle,
where each time point $t$ is mapped to an angle $\theta(t) = 2\pi t/\rho$.
The natural distance on the circle between two time points
$(t,t')$ is given by the chord length of the corresponding points
on the unit circle at angles $(2\pi t/\rho, \;  2\pi t'/\rho)$:
\begin{equation}
  \Delta_{\text{per}}(t,t') = 2\,\abs*{\sin\!\left(\frac{\pi}{\rho}\left|t-t'\right|\right)}.
\end{equation}
This mapping ensures periodic invariance, i.e.,
$k(t,t') = k(t+m\rho,\,t'+n\rho)$ for all integers $m,n \in \mathbb{Z}$.
Using the distance $\Delta_{\text{per}}$ instead of $\Delta_{\text{eu}}$ in the \gpSe{}
kernel~\cref{eq:cov-se} yields the \gpPer{} kernel:
\begin{equation}
  k_{\pPer}(t,t')
  = \sigma^2 \exp\left( -\frac{1}{2\ell^2}{\left(\Delta_{\text{per}}(t,t')\right)^2} \right)
  = \sigma^2
    \exp\left( -\frac{2}{\ell^2} \sin^2\left( \displaystyle\frac{\pi}{\rho} \left|t - t'\right| \right) \right).
\end{equation}
AutoGP's Bayesian inference procedure discovers likely values of the
parameters $(\ell, \rho, \sigma)$ as well as how the \gpPer{} kernel enters
the overall model through appropriate composition operators in the kernel grammar.

\subsection{Extracting Periodicity}
\label{sec:seasonality-extract}

Periodic structure is extracted
from a Gaussian process model structure $m$
using the $\operatorname{split}[m]$ procedure,
shown in \cref{fig:split-periodic}.
It decomposes $m$ into a sum of two subexpressions
$(m_{\rm per}, m_{\rm non})$, where
the former captures the periodic structure
and the latter captures all remaining nonperiodic structure.
The operator distributes over the kernel's sum-of-products expansion,
which ensures that for composite structures like
$(\mathrm{LIN}+\mathrm{PER}) \times \mathrm{GE}$, the resulting cross-term
$\mathrm{PER} \times \mathrm{GE}$ is attributed to the seasonal
component while $\mathrm{LIN} \times \mathrm{GE}$ is not.

\begin{figure}[!h]
\centering
\setlength{\abovedisplayskip}{3pt}
\setlength{\belowdisplayskip}{3pt}

\newcommand{\Zero}{\mathbf{0}}
\begin{nicebox}
\begin{equation*}
\begin{array}[t]{@{}l@{\qquad}l@{\;}l@{\;}l@{}}
(m_{\mathrm{per}},m_{\mathrm{non}}) \defas \operatorname{split}[m]
\\[10pt]
\begin{array}[t]{@{}l}
\operatorname{split}[\mathrm{PER}] \defas (\mathrm{PER},\Zero)
\\
\operatorname{split}[b] \defas (\Zero,b), \; b \neq \mathrm{PER}
\end{array}
&
\begin{array}[t]{@{}l@{}}
\operatorname{split}[m_1+m_2]
\defas
\\
\qquad \text{let }\operatorname{split}[m_1]\gets(a_1,c_1)
\\
\qquad \text{let }\operatorname{split}[m_2]\gets(a_2,c_2)
\\
\qquad \text{in }(a_1 + a_2, c_1 + c_2)
\end{array}
&
\begin{array}[t]{@{}l@{}}
\operatorname{split}[m_1\times m_2]
\defas
\\
\qquad \text{let }\operatorname{split}[m_1]\gets(a_1,c_1)
\\
\qquad \text{let }\operatorname{split}[m_2]\gets(a_2,c_2)
\\
\qquad \text{in }\bigl(a_1\times a_2 + a_1 \times c_2 + c_1 \times a_2, c_1\times c_2\bigr)
\end{array}
&
\end{array}
\end{equation*}
\end{nicebox}
\caption{Recursive definition of $\operatorname{split}[m]$,
which decomposes a covariance kernel expression $m$ into a pair $(m_{\rm
per}, m_{\rm non})$ of kernels, where $m_{\rm per}$ contains the periodic
structure and $m_{\rm non}$ contains the nonperiodic structure.}
\label{fig:split-periodic}
\end{figure}

The decomposition procedure in \cref{fig:split-periodic} ensures
that the covariance kernel $k_m$ for $m$
satisfies $k_m = k_{\rm per} + k_{\rm non}$,
which gives an additive decomposition of the latent process
$X$ into seasonal ($S$) and nonseasonal ($U$) components:
\begin{gather}
  X(t) = S(t) + U(t), \qquad
  S \sim \mathrm{GP}\big(0, k_m^{\mathrm{per}}\big), \quad
  U \sim \mathrm{GP}\big(0, k_m^{\mathrm{non}}\big).
  \label{eq:autogp-xsu-decomposition}
\end{gather}
For time points $\bt\defas (t_1,\dots,t_n)$,
we will momentarily write the random variables
$\mathbf{U}\defas U(\bt)$,
$\mathbf{S}\defas S(\bt)$,
$\mathbf{X}\defas X(\bt)$,
$\mathbf{Y}\defas Y(\bt)$,
$K^{\rm per}_m \defas k_{\rm per}(\bt, \bt)$, and
$K^{\rm non}_m \defas k_{\rm non}(\bt, \bt)$.
The prior distribution is given by
\begin{gather}
  \begin{bmatrix}
    \mathbf{S} \\[2pt] \mathbf{U} \\[2pt] \mathbf{Y}
  \end{bmatrix}
  \sim
  \mathcal{N}\!\left(
  \mathbf{0},\;
  \begin{bmatrix}
    K_m^{\mathrm{per}} & 0 & K_m^{\mathrm{per}} \\[2pt]
    0 & K_m^{\mathrm{non}} & K_m^{\mathrm{non}} \\[2pt]
    K_m^{\mathrm{per}} & K_m^{\mathrm{non}} & K_m^{\mathrm{per}}+K_m^{\mathrm{non}}+\eta I
  \end{bmatrix}
  \right),
  \label{eq:prior-block}
\end{gather}
where, as above, we have a priori independence of $\mathbf{S}$ and $\mathbf{U}$
so that $\mathrm{Cov}(\mathbf{S},\mathbf{U})=0$.
Recall that for any Gaussian vector
\begin{equation}
\begin{bmatrix} \mathbf{Z}_1 \\ \mathbf{Z}_2 \end{bmatrix}
\sim \mathcal{N}\left( \begin{bmatrix} \boldsymbol{\mu}_1 \\ \boldsymbol{\mu}_2 \end{bmatrix},
  \begin{bmatrix}
    \boldsymbol{\Sigma}_{11} & \boldsymbol{\Sigma}_{12} \\
    \boldsymbol{\Sigma}_{21} & \boldsymbol{\Sigma}_{22}
    \end{bmatrix} \right),
\end{equation}
the conditional distribution of $\mathbf{Z}_1$ given $\mathbf{Z}_2$
is also Gaussian, with mean and variance given by
\begin{align}
  \Expect{\mathbf{Z}_1 \mid \mathbf{Z}_2} = \boldsymbol{\mu}_1 + \boldsymbol{\Sigma}_{12} \boldsymbol{\Sigma}_{22}^{-1} (\mathbf{Z}_2 - \boldsymbol{\mu}_2),
  &&
  \Var{\mathbf{Z}_1 \mid \mathbf{Z}_2} = \boldsymbol{\Sigma}_{11} - \boldsymbol{\Sigma}_{12} \boldsymbol{\Sigma}_{22}^{-1} \boldsymbol{\Sigma}_{21}.
\end{align}
Using this result and denoting $K \defas K_m(\bt,\bt)$ yields
the posterior over the seasonal and nonseasonal components:
\begin{align}
  \Expect{\mathbf{S} \,\middle|\, \mathbf{Y},m,\theta,\eta}
    &= K_m^{\mathrm{per}}\, (K+\eta I)^{-1}\mathbf{Y}
  &
  \Expect{\mathbf{U} \,\middle|\, \mathbf{Y},m,\theta,\eta}
    &= K_m^{\mathrm{non}}\, (K+\eta I)^{-1}\mathbf{Y}
  \label{eq:train-means}
  \\[4pt]
  \Var{\mathbf{S} \,\middle|\, \mathbf{Y},m,\theta,\eta}
    &= K_m^{\mathrm{per}} - K_m^{\mathrm{per}}\, (K+\eta I)^{-1} K_m^{\mathrm{per}}
  \label{eq:train-vars}
  &
  \Var{\mathbf{U} \,\middle|\, \mathbf{Y},m,\theta,\eta}
    &= K_m^{\mathrm{non}} - K_m^{\mathrm{non}}\, (K+\eta I)^{-1} K_m^{\mathrm{non}}
  \\[2pt]
  \mathrm{Cov}\left(\mathbf{S},\mathbf{U} \,\middle|\, \mathbf{Y},m,\theta,\eta\right)
    &= -\,K_m^{\mathrm{per}}\, (K+\eta I)^{-1} K_m^{\mathrm{non}}.
  \label{eq:train-cross}
\end{align}
The last term follows from the off-diagonal block of the
joint posterior covariance matrix
$\Var{\mathbf{S}, \mathbf{U} \mid \mathbf{Y}}$.
As off-diagonal block $\mathrm{Cov}(\mathbf{S}, \mathbf{U} \mid \mathbf{Y})$
is generally \textit{nonzero}, the two components
compete to explain the variance.
Adding the means gives the overall posterior mean
$\Expect{\mathbf{X}\mid\mathbf{Y}} = K(K+\eta I)^{-1}\mathbf{Y}$,
and the variances obey
\begin{gather}
  \Var{\mathbf{X} \mid \mathbf{Y}}
  =\Var{\mathbf{S} \mid \mathbf{Y}}
  +\Var{\mathbf{U} \mid \mathbf{Y}}
  +    \mathrm{Cov}(\mathbf{S},\mathbf{U}\mid\mathbf{Y})
  +    \mathrm{Cov}(\mathbf{U},\mathbf{S}\mid\mathbf{Y}),
\end{gather}
which equals the standard GP conditional variance for $K$.
Thus, we have established the following result.

\begin{proposition}[Periodic Posterior Decomposition]
\label{Prop:PostComponDecomp}
For time points $\bt' \defas (t'_1, \dots, t'_{n'})$,
the posterior distribution of the periodic component
$S(\bt') \defas \left(S(t'_1),\dots, S(t'_{n'})\right)$
identified using the $\operatorname{split}$ operator,
given the observation $Y(\bt) = y(\bt)$
and latent variables $(m, \theta, \eta)$ satisfies
\begin{align}
S(\bt') \mid m, \theta, \eta, Y(\bt) = y(\bt)
  &\sim \mathcal{N}\left(
    \mu_{\mathrm{per}}^{(m,\theta,\eta)}(\bt'),
    \Sigma_{\mathrm{per}}^{(m,\theta,\eta)}(\bt',\bt')
    \right),
  \\
\shortintertext{where}
\mu_{\mathrm{per}}^{(m,\theta,\eta)}(\bt')
  &\defas K_m^{\mathrm{per}}(\bt',\bt)
  \left[K_m(\bt,\bt) + \eta I\right]^{-1} \mathbf{y}(\bt)
  \label{eq:seasonal-mean},
  \\
  \Sigma_{\mathrm{per}}^{(m,\theta,\eta)}(\bt',\bt')
  &\defas K_m^{\mathrm{per}}(\bt',\bt')
  - K_m^{\mathrm{per}}(\bt',\bt)
  \left[K_m(\bt,\bt) + \eta I\right]^{-1}
  K_m^{\mathrm{per}}(\bt,\bt').
  \label{eq:seasonal-cov}
\end{align}
Using properly weighted posterior samples
$\set*{\left(w^{(i)}, \left(m^{(i)}, \theta^{(i)}, \eta^{(i)}\right)\right)}_{i=1}^M$,
the posterior law of the seasonal component is approximated
by a finite mixture that averages over posterior uncertainty in model structure,
kernel parameters, and observation noise variance:
\begin{equation}
\mathcal{L}\left( \mathbf{S}(\bt') \mid Y(\bt) = y(\bt) \right)
  \approx \sum_{i=1}^M \tilde w^{(i)}
  \mathcal{N}\left(
    \mu_{\mathrm{per}}^{(i)}(\bt'),
    \Sigma_{\mathrm{per}}^{(i)}(\bt',\bt')
    \right),
  \qquad \tilde{w}^{(i)} \defas w^{(i)} \big/ \sum_{j=1}^M w^{(j)}
    \quad (1 \le i \le M),
  \label{eq:bma-mixture}
\end{equation}
where $\mu_{\mathrm{per}}^{(i)}(\bt')$ and
$\Sigma_{\mathrm{per}}^{(i)}(\bt',\bt')$
are the mean and covariance of the periodic component for the
$i$\textsuperscript{th} posterior draw $\left(m^{(i)}, \theta^{(i)}, \eta^{(i)}\right)$,
as defined in \cref{eq:seasonal-mean,eq:seasonal-cov}.
\end{proposition}

\Cref{Prop:PostComponDecomp}, whose proof is in \cref{appx:gp-proof},
characterizes seasonality as a posterior distribution over latent
seasonal paths.
This probabilistic representation is what we mean by
\emph{probabilistic seasonality}.

\subsection{Identifying the Seasonal Component}
\label{sec:seasonality-identify}

\Cref{Prop:PostComponDecomp} gives the posterior distribution of the raw
periodic component $S(\bt')$ at an arbitrary collection of time points
$\bt'$.
To interpret this component as an econometric
seasonal factor, however, we must impose a normalization.

\begin{figure}[!h]
\centering
\includegraphics[width=\textwidth]{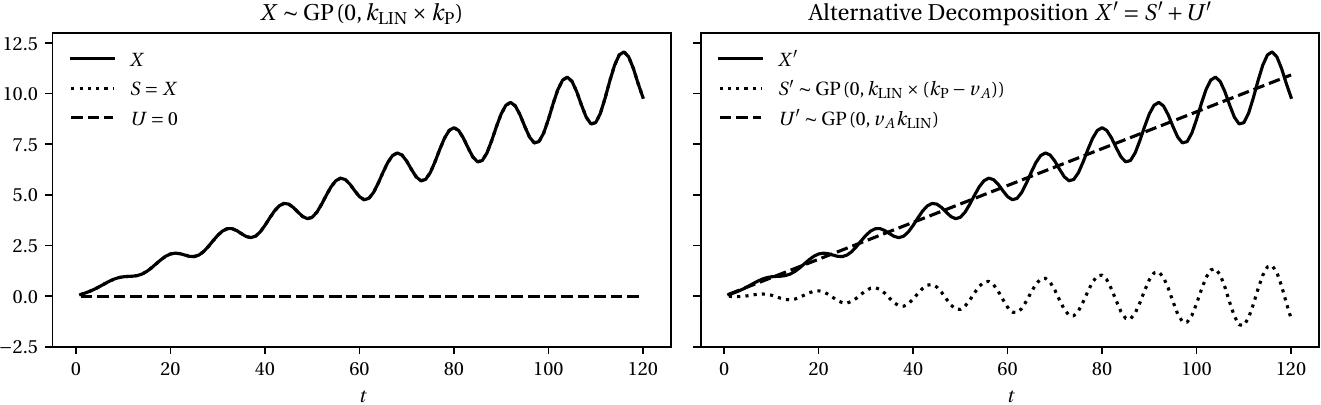}
\caption{Seasonal components are not identified.
The left panel shows a model with a multiplicative seasonal component.
The right panel shows a distributionally equivalent model
with a nonzero nonseasonal component.}
\label{fig:identification}
\end{figure}

It is well known in the seasonal
adjustment literature \citep{Watson:1987,harvey1989forecasting},
that the decomposition
$Y(t)=S(t)+U(t)+\varepsilon(t)$
of a series into seasonal, nonseasonal, and residual components
is not uniquely identified.
For example, the likelihood is invariant to the transformation
$S'(t)=S(t)+a$ and $U'(t)=U(t)-a$ for any constant $a$.
Such identification issues appear in the Gaussian process model family from
\cref{sec:discovery} as follows.
Let $k_\pLin$ and $k_\pPer$ denote a linear and 12-month periodic covariance
kernel, respectively, and consider a process
$X \sim \mathrm{GP}\left(0, k_\pLin \gpMul k_\pPer\right)$
with a multiplicative periodic component $S$ and a zero nonperiodic component $U$.
This time series admits an alternative decomposition as a sum of
a seasonal and nonseasonal Gaussian processes:
\begin{align}
X \overset{\mathrm{D}}{=} X ' \defas S' + U',
\quad
S' \sim \mathrm{GP}\left(0, k_\mathrm{LIN}\times(k_\mathrm{P}-v_A)\right),
\quad
U' \sim \mathrm{GP}\left(0, v_A k_\mathrm{LIN}\right),
\quad
S' \perp U',
\end{align}
where $v_A \defas \mathrm{Var}\left(\sum_{t=1}^{12} W(t)/12\right)$ for $W \sim \mathrm{GP}(0, k_\pPer)$.
It follows that the $\operatorname{split}$ procedure from \cref{fig:split-periodic}
returns a different decomposition when applied to $X$ or $X'$, even though they
are distributionally equivalent on the monthly grid.
\Cref{fig:identification} demonstrates this effect.
Note that the alternative representation $X'$ is not contained in the
prior~\crefrange{eq:hbm-base}{eq:hbm-m}, while the original representation $X$ is,
which underscores the role of the kernel language in identification.

As a unique decomposition is impossible, we follow standard econometric
conventions by imposing a zero-mean normalization over a fixed reference
window of time points.
This step removes the additive level ambiguity by assigning the average
seasonal level on that window to the nonseasonal component.
The remaining allocation of persistent variation is determined by the
covariance kernel representation and split rule.

Our normalization procedure operates as follows.
Let $\br \defas (r_1,\ldots,r_q)$ denote a finite sequence of time points
on which the seasonal component is to be normalized (called the normalization window).
For in-sample seasonal adjustment, $\br = \bt$ is the list of time points in the observed dataset.
For forecast evaluation, $\br = \bt'$ is the list of time points in the forecast window.
Define the
$P_q \defas \frac{1}{q}\mathbf 1_q\mathbf 1_q^\top$
to be the projection operator on the constant vector
and $C_q \defas I_q-P_q$, to be the centering operator,
respectively.
Thus, for any vector $v\in\mathbb R^q$, $C_qv$ subtracts the sample average
of $v$ over the length-$q$ window $\br$.
The \textit{normalized seasonal component} on $\br$ is the random vector
\begin{align}
\widetilde S_\br  \defas C_q S(\br),
&&
[\widetilde S_\br ]_j = S(r_j)-\frac{1}{q}\sum_{\ell=1}^q S(r_\ell)
\quad (1 \le j \le q).
\end{align}
%
By construction, $\mathbf 1_q^\top \widetilde S_\br =0$,
so the normalized seasonal component has zero
average over the normalization window.
The removed level component is reassigned to the
\textit{normalized nonseasonal component} on $\br$:
\begin{align}
\widetilde U_\br \defas U(\br)+P_qS(\br),
&&
[\widetilde U_\br]_j
=
U(r_j)+\frac{1}{q}\sum_{\ell=1}^q S(r_\ell)
\quad (1 \le j \le q).
\end{align}
These transformations preserve the latent Gaussian process $X$ on $\br$,
by reallocating the removed level component from the seasonal to the
nonseasonal component:
\begin{equation}
\widetilde S_\br +\widetilde U_\br
=
C_qS(\br)+U(\br)+P_qS(\br)
=
S(\br)+U(\br)
=
X(\br).
\end{equation}

Normalization only removes the constant level
of the extracted seasonal component over the window,
but it does not remove time variation in the periodic structure itself.
For example, composite terms such as
$\mathrm{PER}\times\mathrm{GE}$ or $\mathrm{PER}\times\mathrm{LIN}$ still
remain part of the normalized seasonal component $\widetilde{S}_{\br}$,
which allows periodic structure to evolve over time
through changes in local smoothness or amplitude.

\subsection{Posterior Law of Seasonal Component}
\label{sec:seasonality-law}

In our probabilistic seasonality framework, the normalized seasonal
component $\widetilde S_\br$ is a bona fide random vector.
Because the normalization step
applies a linear transformation of the raw
seasonal component $S(\br)$, the posterior distribution
from \cref{Prop:PostComponDecomp} can be pushed forward directly.
Therefore, seasonal normalization, uncertainty quantification,
and seasonal adjustment all live within the same inferential framework.
Recall that for a fixed posterior draw $(m^{(i)},\theta^{(i)},\eta^{(i)})$, the
raw seasonal component satisfies
\begin{equation}
S(\br) \,\big\vert\, m^{(i)},\theta^{(i)},\eta^{(i)},Y(\bt)=y(\bt)
\sim
\mathcal N_q\!\left(
  \mu_{\mathrm{per}}^{(i)}(\br),
  \Sigma_{\mathrm{per}}^{(i)}(\br,\br)
  \right).
\end{equation}
Therefore the normalized seasonal component has the Gaussian law
\begin{equation}
\widetilde S_\br \,\big\vert\, m^{(i)},\theta^{(i)},\eta^{(i)},Y(\bt)=y(\bt)
\sim \mathcal N_q\!\left(
  C_q\mu_{\mathrm{per}}^{(i)}(\br),
  C_q\Sigma_{\mathrm{per}}^{(i)}(\br,\br)C_q^\top
  \right),
\end{equation}
which is supported on the zero-sum subspace
$\set{ v \in \mathbb{R}^q \mid \mathbf{1}_q^\top v = 0 }$.
Marginalizing over the hidden structure and parameters,
the posterior distribution of $\widetilde S_\br$
can be approximated using the particle collection
\begin{equation}
\mathcal L\!\left(\widetilde S_\br \mid Y(\bt)=y(\bt) \right)
\approx   \sum_{i=1}^M
    \widetilde w^{(i)}\,
    \mathcal N_q\!\left(
      C_q\mu_{\mathrm{per}}^{(i)}(\br),
      C_q\Sigma_{\mathrm{per}}^{(i)}(\br,\br)C_q^\top
    \right),
\qquad
\tilde{w}^{(i)} \defas w^{(i)} \big/ \sum_{j=1}^M w^{(j)} \quad (1 \le i \le M).
\end{equation}
The posterior mean is therefore approximated as
\begin{equation}
\widehat s_\br \defas \Expect{\widetilde S_\br \mid Y(\bt)=y(\bt)}
  \approx \sum_{i=1}^M \widetilde w^{(i)} C_q\mu_{\mathrm{per}}^{(i)}(\br).
\end{equation}
When the normalization window satisfies $\br=\bt$,
we obtain in-sample seasonally adjusted series
\begin{equation}
\widehat y^{SA}(\bt) \defas y(\bt)-\widehat s_\bt.
\end{equation}

\subsection{Seasonal Queries}
\label{sec:seasonality-queries}

The posterior law of $\widetilde S_\br$ given $\set{Y(\bt) = y(\bt)}$
supports several queries that are either ill-posed or difficult to answer
using standard seasonal-adjustment procedures.
For compactness, we define
$\widetilde\mu^{(i)}(\br) \defas C_q\mu_{\mathrm{per}}^{(i)}(\br)$
and
$\widetilde\Sigma^{(i)}(\br) \defas C_q\Sigma_{\mathrm{per}}^{(i)}(\br,\br)C_q^\top$
to be the normalized posterior mean and covariance of $\widetilde S_\br$,
respectively, according to particle $i \in \set{1,\dots,M}$.
The following seasonality queries are solved using the inferred posterior law:
no auxiliary statistical tests or per-query re-estimation procedures are required.

\paragraph{Existence of Seasonality.}
The posterior probability that seasonality is present is the posterior
probability that the sampled kernel structure has a nonzero periodic component
after applying the $\operatorname{split}$ operator from \cref{fig:split-periodic}.
Let $(m_{\mathrm{per}}, m_{\mathrm{non}}) \defas \operatorname{split}(m)$
for a kernel structure $m$; the posterior probability of periodic
structure is approximated as
\begin{equation}
\Pr\!\left(m_{\mathrm{per}}\neq 0\mid Y(\bt)=y(\bt)\right)
\approx
\sum_{i=1}^M \widetilde w^{(i)}  \mathbf 1\!\set*{m_{\mathrm{per}}^{(i)}\neq 0}.
\label{eq:query-prob-seasonality}
\end{equation}

\paragraph{Strength of Seasonality.}
The seasonal magnitude
over a normalization window $\br$ within a particle $i$ is the random variable
\begin{align}
A^{(i)}(\br) \defas \frac{1}{\sqrt{q}} \left\lVert Z^{(i)} \right\rVert_2,
  && Z^{(i)} \overset{\mathcal L}{=} \widetilde{S}_\br \mid m^{(i)}, \theta^{(i)}, \eta^{(i)}, Y(\bt) = y(\bt),
  && (1 \le i \le M).
\end{align}
This quantity is nonnegative and in the same units as the modeled series.
All summaries, such as the mean, median, or tail probabilities, of the
seasonal magnitude variable are estimated from a coherent posterior
distribution using the collection of $M$ particles.

\paragraph{Sign and Timing.}
The posterior probability that the seasonal effect is
positive at time $r_j$
in the normalization window $\br=(r_1,\ldots,r_q)$
is approximated as
\begin{align}
\Pr\!\left(
  \left[\widetilde S_\br\right]_j>0\mid Y(\bt)=y(\bt)\right)
  \approx \sum_{i=1}^M
    \mathbf{1}\left[m^{(i)}_{\mathrm{per}} \ne 0\right]
    \widetilde w^{(i)}
    \Phi\!\left( \frac{[\widetilde\mu^{(i)}(\br)]_j} {\sqrt{[\widetilde\Sigma^{(i)}(\br)]_{jj}}}
\right).
\end{align}
This probability quantifies the posterior confidence that seasonality raises
the observed time series relative to the nonseasonal component at
time $r_j$.

\paragraph{Model Disagreement.}
The posterior covariance of the normalized seasonal component
decomposes into
within-particle uncertainty and between-particle disagreement,
by the law of total variance:
\begin{align}
\mbox{Overall uncertainty}:\quad &\Var{\widetilde S_\br\mid Y(\bt)=y(\bt)} \approx W_\br+B_\br \\
\mbox{Within particle uncertainty}:\quad  &W_\br \defas \textstyle\sum_{i=1}^M \widetilde w^{(i)} \widetilde\Sigma^{(i)}(\br) \\
\mbox{Between particle uncertainty}:\quad &B_\br \defas \textstyle\sum_{i=1}^M \widetilde w^{(i)} \left(\widetilde\mu^{(i)}(\br)-\widehat s(\br)\right) \left(\widetilde\mu^{(i)}(\br)-\widehat s(\br)\right)^\top.
\end{align}
Large diagonal entries of $B_\br$ indicate time points where posterior particles
disagree about the seasonal effect.
The scalar $\operatorname{tr}(B_\br)/q$
summarizes the average between-particle disagreement over the window.

\paragraph{Revision Risk.}

Standard diagnostics for the revision history of seasonal adjustments
operate retrospectively, by comparing an initial adjustment with new
estimates obtained \textit{after} additional observations have become
available \citep{MonsellFindley1984}.
Using \SGP{}, we instead characterize a full posterior
distribution over future revisions to a current adjustment,
\textit{before} any additional data is observed.

Let $\bt \defas (t_1,\ldots,t_n)$ be the current observed time points
with observed data $Y(\bt)=y(\bt)$
and $\bt' \defas (t_{n+1},\ldots,t_{n+h})$ be $h > 0$ future points.
For a possible realization $y(\bt')$ of the future data $Y(\bt')$,
define $\widehat s_\br(y(\bt'))$ as the posterior mean
of the normalized seasonal component $\widetilde S_{\br}$
conditioned on all $n+h$ observations $Y(\bt, \bt')=y(\bt,\bt')$..
The \textit{revision} in the estimated seasonal adjustment at time $r_j$
(typically $r_j = t_j$)
is a deterministic function of the new data:
\begin{align}
R_{j}(y(\bt')) \defas \left[\widehat s_\br(y(\bt'))\right]_j - \left[\widehat s_\br\right]_j
&& (1 \le j \le q).
\label{eq:defn-revision}
\end{align}
With probabilistic seasonality, we can quantify the full predictive
distribution of the future revisions $R_{j}$ conditioned on
the current data $Y(\bt) = y(\bt)$, but \textit{before} observing
the new data $Y(\bt') = y(\bt')$.
The posterior predictive distribution of $Y(\bt')$ is therefore
\begin{align}
\mathcal{L}\left( Y\left(\bt'\right) \mid Y(\bt) = y(\bt) \right)
  \approx \sum_{i=1}^{M}\widetilde w^{(i)} \mathcal{N}\left( \mu_{\rm post}^{(i)}(\bt'), \Sigma_{\rm post}^{(i)}(\bt',\bt') + \eta^{(i)} I_{h} \right),
\label{eq:malcontentedly}
\end{align}
where $\mu_{\rm post}^{(i)}(\bt')$ and
$\Sigma_{\rm post}^{(i)}(\bt',\bt')$ are the particle-specific posterior mean
and covariance of the latent process $X(\bt')$ conditioned on $Y(\bt)=y(\bt)$
(\cref{appx:gp-model}).
This distribution induces a distribution over the future revisions:
\begin{align}
\Pr\!\left(R_{j}\left( Y(\bt')\right) \in A\mid Y(\bt)=y(\bt)\right)
   &= \Expect{\mathbf{1} \set{ R_{j}(Y(\bt')) \in A} \mid Y(\bt) = y(\bt)}
  && (A \subset \mathbb{R}\;\; \mbox{measurable}),
\label{eq:revision-risk-distribution}
\end{align}
which can be used to directly quantify the expected
magnitude
$\Expect{\abs*{R_{j}(Y(\bt'))} \;\middle|\; Y(\bt)=y(\bt)}$
as well as other statistics related to future revisions.

\section{A Simulation Study}
\label{sec:simulation}

We next assess the efficacy of \SGP{} through a simulation exercise.
Because seasonality is a latent component with no universally-accepted
definition, the ``true'' seasonal component of a time series is not
directly observable from data.
For this reason, simulations provide an environment in which the underlying
seasonal structure is known by construction and methodologies
can be assessed transparently~\citep{GhyselsPerron1990SeasonalUnitRoot}.

\begin{figure}[htb]
\centering
\includegraphics[width=.75\textwidth]{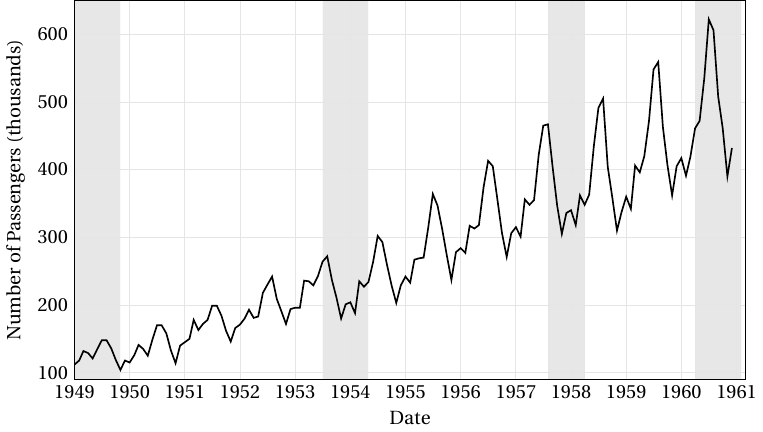}
\caption{Passengers on International Flights 1949--1960.}
\label{fig:airline}
\end{figure}

The simulated data are designed to capture characteristics common
to seasonal economic time series.
To that end, consider \cref{fig:airline}, which
plots the number of passengers per month on international flights on
U.S. air carriers from January 1949 through December 1960.
The grey shaded bars indicate NBER recessions.
This time series is representative in the sense that it
displays three characteristics common to economic data:
\begin{enumerate*}[label=(\roman*)]
\item a saw-tooth pattern due to seasonality as air travel peaks in summer months and
troughs in the winter;
\item an upward trend that either slows or declines with recessions; and
\item an increase in peak-to-trough dynamics over time.
\end{enumerate*}

The dataset in \cref{fig:airline}, originally sourced from
the FAA Statistical Handbook of Civil Aviation and first appearing in
\citet[Table C.10]{Brown:TimeSeries:1963}, has become a classic benchmark for
assessing seasonality, extensively analyzed in seminal textbooks such
as \citet{BoxJenkins:TimeSeries:1976} and used as a baseline specification
for the X-13-ARIMA-SEATS algorithm (referred to as the ``airline model'').

\subsection{Methodology}
\label{sec:simulation-methodology}

\subsubsection{Data Generating Process}
\label{sec:simulation-methodology-dgp}

Consider a discrete-time stochastic process given by
\begin{equation}
Y(t) \defas
    a_0
    + a_1 \cdot t
    + (a_2 + a_4 \cdot t) \sin \left(\omega \cdot t\right)
    + (a_3 + a_5 \cdot t) \cos \left(\omega \cdot t\right)
    + \varepsilon(t)
    \label{eq:model-growing-sin}
\end{equation}
where $t \in \set{1,2,3,\dots,T}$ is a discrete time index (e.g., months);
$\omega = 2 \pi / 12$ is the angular frequency corresponding to a 12-month
(annual) seasonal cycle; and
$\varepsilon(t) \sim \mathcal{N}(0, \sigma^2_\varepsilon)$
is a Gaussian error term (i.i.d.).
This model captures essential features of the
economic time series in \cref{fig:airline} through a
linear time trend ($a_1 \cdot t$),
seasonal terms ($a_2 \cdot \sin \left(\omega t\right)$, $a_3 \cdot \cos \left(\omega t \right)$)
whose amplitudes grow linearly in time ($a_4 \cdot t$, $a_5 \cdot t$),
which allows for increasing peak-to-trough dynamics.
To ensure a constant phase, the parameters satisfy
$a_2a_5 = a_3a_4$, reducing the effective number of parameters to five.
The seasonal component can be expressed as a single sinusoid
\begin{equation}
    s(t) \defas r(t) \sin \left(\omega \cdot t + \phi\right), \quad
    r(t) \defas \sqrt{(a_2 + a_4 \cdot t)^2 + (a_3 + a_5 \cdot t)^2}, \quad
    \phi \defas \arctan2(a_3, a_2),
    \label{eqn:SeasonalGrowingSineModel}
\end{equation}
where $r(t)$ is the amplitude, which grows linearly with time, and
$\phi$ is the constant phase.
This model specification is discussed in \citet[Chapter 4]{Brown:TimeSeries:1963}.

\subsubsection{Simulation}
\label{sec:simulation-methodology-simulation}

The model~\cref{eq:model-growing-sin} is fit to the Air Passengers data of
\cref{fig:airline} using ordinary least squares (OLS).
The fitted parameters are
$\hat{a}_0=87.82$,
$\hat{a}_1=2.65$,
$\hat{a}_2=-0.24$,
$\hat{a}_3=-0.57$,
$\hat{a}_4=-0.25$,
$\hat{a}_5=-0.58$, and
$\hat{\sigma}_\varepsilon = 26.90$,
with an adjusted R-squared of 0.95.
Note that our OLS fit does not impose the constant-phase constraint
$a_2a_5 = a_3a_4$, but the
phase change implied by the fitted parameters is negligible
($\hat a_2 \hat a_5 \approx 0.1392$ and $\hat a_3 \hat a_4 \approx 0.1425$).
This fitted model is used as the true data generating process for the
simulation study.
For simulation length $T$, we generate $T$ observations
$y(\bt) \defas (y(1), \dots, y(T))$ to serve as the observed data
at time points $\bt \defas (1,\dots,T)$, followed
by 12 observations
$y(\bt') \defas (y(T+1), \dots, y(T+12))$
for out-of-sample forecasting at time points $\bt' \defas (T+1, \dots, T+12)$.
We examine sample sizes of $T \in \set{36, 72, 144}$, corresponding
to 3, 6, and 12 years of monthly data,
to assess forecasting performance across small to large training samples.
For each sample size of length $T$, we produce $N = 200$ simulated time series,
each consisting of $T + 12$ data points.

\subsubsection{Estimation}
\label{sec:simulation-methodology-estimation}

Each simulated time-series dataset is fit using three competing approaches:
\begin{itemize}[nosep]
\item the correctly specified model of \eqref{eq:model-growing-sin} using OLS (without the
constant-phase constraint $a_2a_5 = a_3a_4$);
\item the AutoGP probabilistic seasonality approach described in \cref{sec:discovery,sec:seasonality}; and
\item a model estimated using the X-13 ARIMA-SEATS (X-13 hereafter) algorithm.
\end{itemize}
Both model parameters \emph{and} model structure must be learned when
implementing probabilistic model discovery and X-13ARIMA-SEATS.
Because OLS already uses the correctly specified model structure, standard
asymptotic arguments apply to OLS estimates, which will serve as the
gold-standard reference benchmark.

The most relevant comparison model is X-13,
which is the current standard in seasonal adjustment developed over many
decades at various statistical agencies and described in
\cite{X13RefManual:2024}.
It is a comprehensive statistical algorithm used by the vast majority of
governments to deseasonalize economic data, including the U.S.~Bureau of
Economic Analysis (BEA), the U.S.~Bureau of Labor Statistics (BLS), Eurostat,
and Statistics Canada.%
\footnote{The BEA uses X-13 to seasonally adjust many of
    its key economic indicators, including Gross Domestic Product and all components,
    Personal Income and Outlays, many international transactions such as trade balances.
    The BLS uses X-13 to seasonally adjust nearly all employment statistics
    (unemployment rate, hours worked), and all price indices (Consumer Price and
    Producer Price Indices).}
Our analysis uses X-13ARIMA-SEATS Version 1.1 Build 61, July 10, 2024.%
\footnote{JDemetra+, developed by the National
    Bank of Belgium (NBB) in collaboration with Eurostat and the European Central Bank
    (ECB), is a Java-based, open-source alternative to X-13.
    Given that we focus on U.S.\ data and that both programs
    implement nearly identical methodologies, we do not include JDemetra+
    in our analysis.}
It is important to note that X-13 is an algorithm in the sense that the user
can specify \textit{automated} responses at all steps based on optimization
routines and well-established statistical thresholds\footnote{%
In the empirical and simulation exercises, our automated X-13 specification is
\texttt{transform{function=auto}},\texttt{regression{aictest=(td easter)}},
	\texttt{automdl{}},
	\texttt{outlier{}}.
Thus X-13 automatically chooses the transformation, tests for trading-day and
	Easter calendar regressors, selects the regARIMA model, detects outliers, and
	then applies the X-11 seasonal-adjustment decomposition.},
which enable us to compare X-13 to \SGP.

\subsubsection{OLS Gold Standard}
\label{sec:simulation-methodology-ols}

OLS serves as a gold-standard baseline because it is given the exact
structural form of the simulation data-generating process~\cref{eq:model-growing-sin},
which is linear in the coefficients $(a_0,\ldots,a_5)$.
The OLS estimates exhibit two sources of uncertainty: innovation noise
uncertainty and parameter uncertainty.
In particular, forecasting the observed series $Y(T+h)$ requires accounting
for a new innovation $\varepsilon(T+h)$, so even the correctly specified
gold-standard forecast has irreducible aleatoric uncertainty.
By contrast, the latent seasonal component $s(T+h)$ is deterministic,
conditional on the regression coefficients, so seasonal forecast error
arises only from parameter uncertainty.
To make these effects explicit, write \cref{eq:model-growing-sin} as
\begin{equation}
Y(t)=\mathbf z_t^\top \beta+\varepsilon(t),
\qquad
\mathbf z_t \defas \left(1,\, t,\, \sin(\omega t),\, \cos(\omega t),\, t\sin(\omega t),\, t\cos(\omega t) \right)^\top,
\quad
\beta \defas (a_0,a_1,a_2,a_3,a_4,a_5)^\top .
\end{equation}
The OLS forecast error for the observed series at horizon $h$ is
\begin{equation}
\widehat Y(T+h) - Y(T+h) = \mathbf z_{T+h}^\top\left(\widehat\beta-\beta\right) - \varepsilon(T+h).
\end{equation}
Thus the forecast error contains both the new innovation
$\varepsilon(T+h)$ and coefficient estimation error $\hat{\beta} - \beta$.
The innovation does not affect the mean forecast of the observed series,
but it does affect the realized forecast error and prediction intervals.
In contrast, the seasonal components and OLS forecast errors are
\begin{align}
s(t) &= \mathbf q_t^\top\beta,
&&
\mathbf{q}_t \defas \left(0,\, 0,\, \sin(\omega t),\, \cos(\omega t),\, t\sin(\omega t),\,t\cos(\omega t) \right)^\top,
\\
\widehat{s}(T+h) - s(T+h) &= \mathbf{q}_{T+h}^\top(\widehat\beta-\beta).
\end{align}
There is no irreducible innovation term in the seasonal forecast error.
OLS thus provides a gold-standard reference point for
assessing the cost of discovering seasonal structure (as in AutoGP and X-13)
rather than being given the seasonality in advance.
Performance close to OLS indicates that little accuracy is lost from structure discovery.

Our simulation study uses $N=200$ datasets for each $T \in \set{36,72,144}$,
which is the smallest number of simulations for which the gold-standard OLS
distribution of RMSE and MIS remained stable.

\subsubsection{Evaluation}
\label{sec:simulation-methodology-evaluation}

For each method, we evaluate forecasts of the held-out observed data $Y(T+h)$ and
a normalized seasonal component
$\widetilde{s}_{\bt'} \defas C_{12}\,s(\bt')$, where the normalization
centers over the forecast window
$\bt' \defas (T+1, \dots, T+12)$ using the centering matrix $C_{12}$.
While this normalization does \emph{not} alter our simulation results in
any substantive manner, it is consistent with the empirical approach of
the next section and serves to remove the level ambiguity inherent in the trend–seasonal
decomposition.
For AutoGP, this normalization is applied draw-by-draw to posterior seasonal
paths before posterior means and quantiles are computed.

For the data target, relative RMSE of a particular method at horizon $h$
is given by
\begin{equation}
\text{Relative RMSE}_{h}
    \defas
    \frac{\text{RMSE}_{h}}{\text{RMSE}{h}^{(\text{OLS})}}-1,
\qquad
\text{RMSE}_{h}
    \defas
    \left[
        \frac{1}{\abs{\mathcal I_{h}}}
        \sum_{i\in\mathcal I_{h}}
            \left(
            \widehat y^{(i)}(T+h)-y^{(i)}(T+h)
            \right)^2
        \right]^{1/2}
\end{equation}
where $\mathcal I_h$ is the set of replications for which the method
produces a forecast at horizon $h$.
OLS and AutoGP produce forecasts for all $200$ replications at each $T$.
X-13 fails to converge for 21 of the 600 simulated series, which
gives 192, 197, and 190 replications for $T=36,72,144$, respectively.
The analogous seasonal RMSE is computed using the normalized target
vector $\widetilde s^{(i)}(\bt')$ and normalized seasonal forecast
$\widehat{\widetilde s}^{(i)}(\bt')$.

We also evaluate the quality of central $(1-\alpha)$ predictive intervals with
the mean interval score (MIS), which jointly evaluates the coverage
and sharpness.
For $\alpha=0.05$, scalar target $z$, and prediction interval $[\ell,u]$, define
\begin{equation}
\mathrm{MIS}_{\alpha}(z;\ell,u)
=
\left(u-\ell\right)
+
\frac{2}{\alpha}
\left(\ell-z\right)^{+}
+
\frac{2}{\alpha}
\left(z-u\right)^{+},
\qquad
x^{+}\defas \max\set{x,0}.
\end{equation}
For the data target we have $z=Y(T+h)$,
and for the seasonal target we have $z=\widetilde s(T+h)$.
X-13 produces predictive intervals for the observed data target but not usable
predictive intervals for the X-11 seasonal component in this exercise; hence
X-13 seasonal MIS is not reported.

Finally, we conduct paired AutoGP--X-13 tests with significance
based on Benjamini--Hochberg (BH) adjusted $q$-values across the twelve
forecast horizons within each target and sample length.
For each horizon $h$, the unit of analysis is the matched pair of per-dataset
losses $L^{(\mathrm{AutoGP})}_{i,h}$ and $L^{(\text{X-13})}_{i,h}$
(squared error for RMSE and interval score for MIS)
computed on the same simulated dataset $i$.
We test the null hypotheses
$H_{0,h}\!:\,\E\!\big[L^{(\mathrm{AutoGP})}_{i,h}-L^{(\text{X-13})}_{i,h}\big]=0$
with a two-sided paired $t$-test over the $N=200$ replications.

\subsection{Results}
\label{sec:simulation-results}

\subsubsection{Data Target}
\label{sec:simulation-results-data}

\begin{figure}[!p]
\centering
\includegraphics[width=\textwidth]{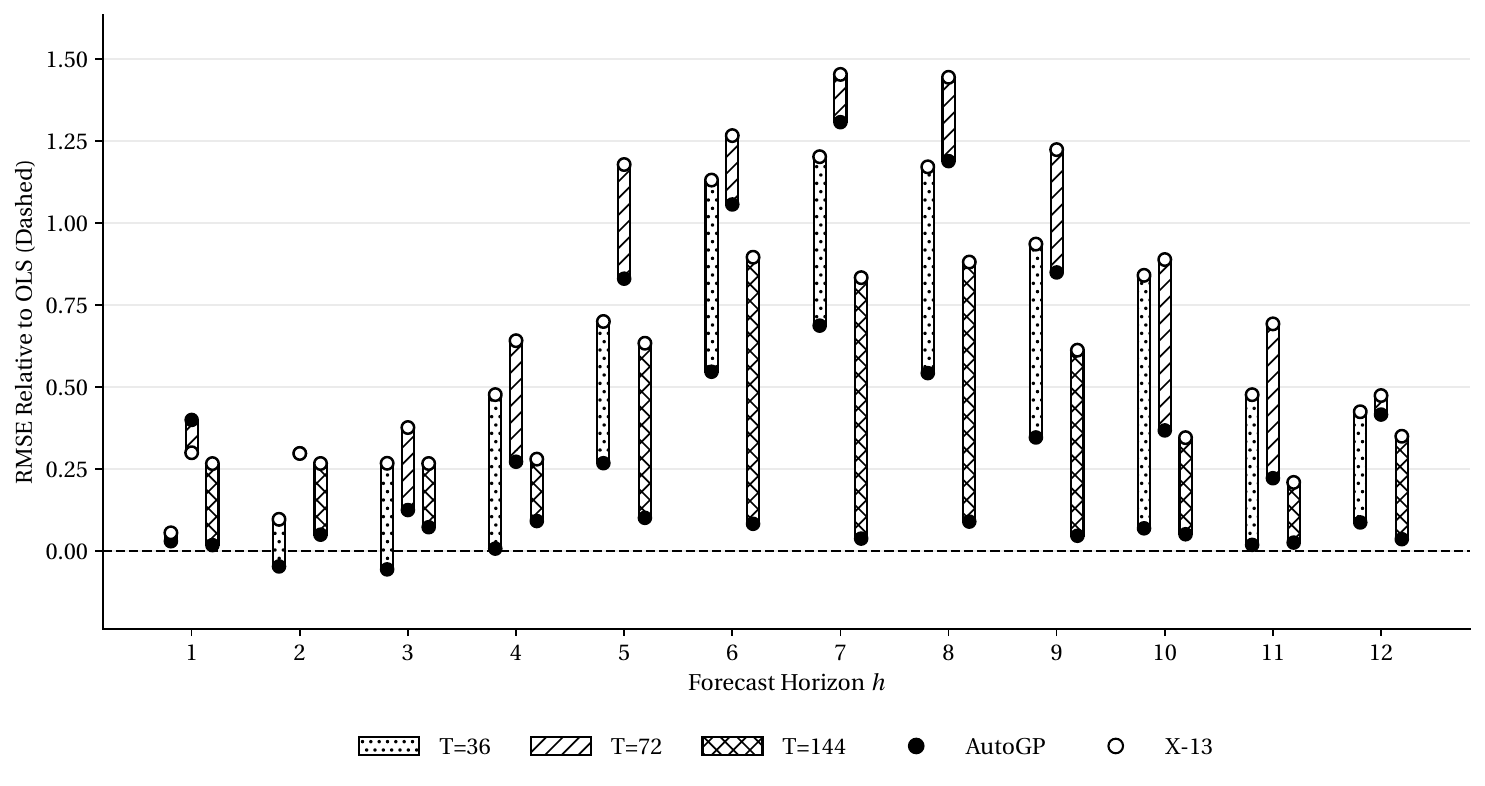}
\caption{Relative RMSE performance for the data target.}
\label{fig:airline-data-rmse}

\includegraphics[width=\textwidth]{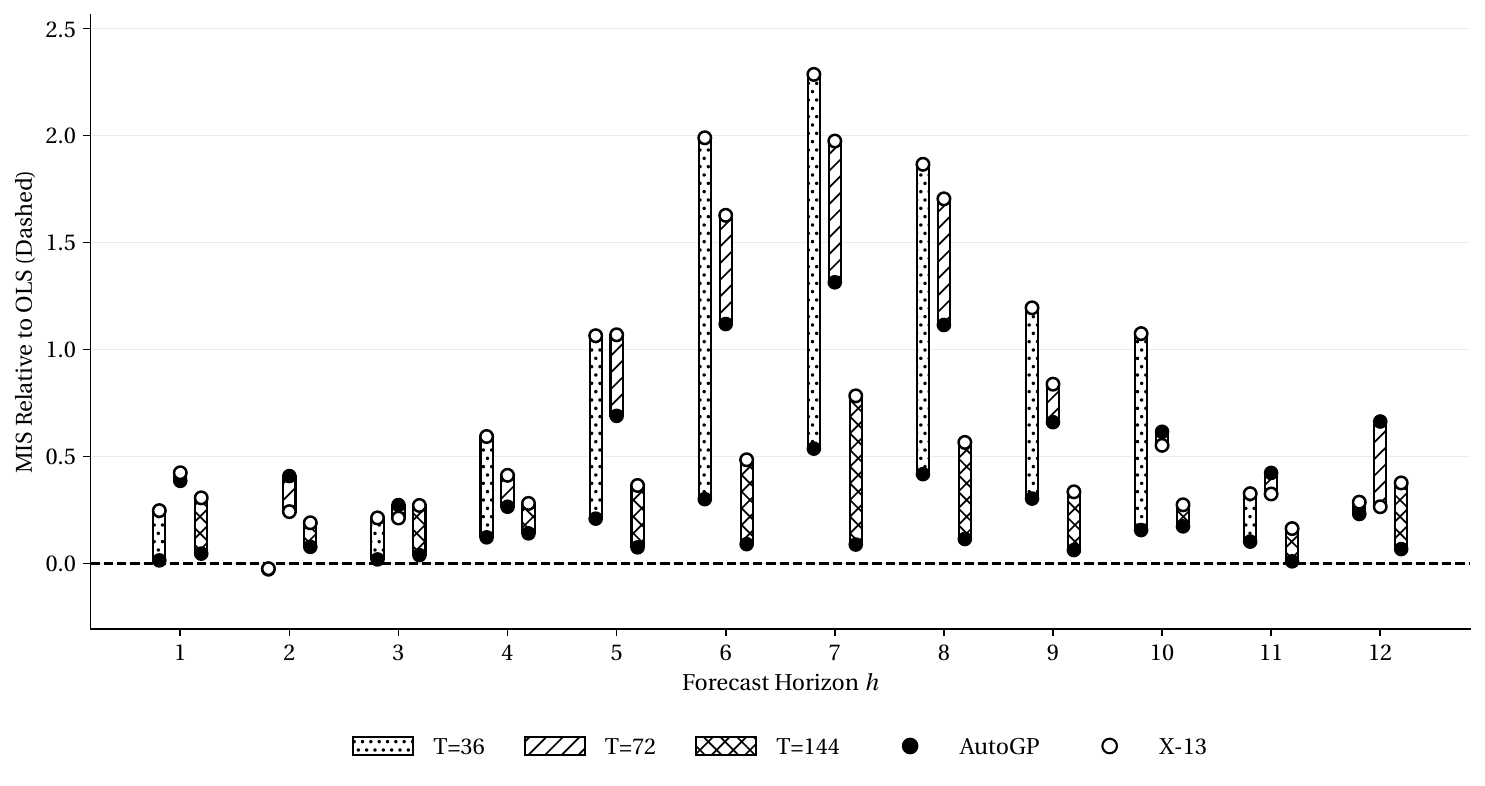}
\caption{Relative MIS performance for the data target.}
\label{fig:airline-data-mis}
\end{figure}

\Cref{fig:airline-data-rmse} displays the relative RMSE
across forecast horizons $h=1, \dots, 12$.
The OLS baseline is centered exactly at zero and values above zero
indicate higher (worse) error than OLS, while values below
indicate lower (better) error.
Circles denote the relative RMSE for the AutoGP estimated models
with training lengths $T=36$ (green), $T=72$ (orange), and $T=144$ (purple).
Squares denote the relative RMSE for the X-13 model.
Vertical bars connecting the X-13 and AutoGP markers highlight
the magnitude and direction of the performance gap between the
two models at each horizon.

\Cref{fig:airline-data-rmse} illustrates that
AutoGP consistently achieves a lower RMSE than the X-13 baseline across nearly
all horizons and sample sizes.
Given a sufficient sample size ($T=144$), AutoGP uniformly dominates, yielding
statistically significant MSE reductions over X-13 across all twelve forecast
horizons ($q < 0.05$).
AutoGP approaches the OLS benchmark for $T=144$, with an RMSE within 5\% at all
horizons.
At the short sample size ($T=36$) AutoGP outperforms
X-13 at all horizons ($h \geq 2$, $q < 0.01$),
and even outperforms the OLS specification at certain horizons.
At the intermediate sample size ($T=72$), the performance gap narrows.
AutoGP achieves statistically significant error reductions at intermediate
horizons ($h \in \set{3,4,5,9,10,11}$), while performing statistically on par
with X-13, except for $h=1$ where X-13 slightly outperforms.

\Cref{fig:airline-data-mis} reports the quality of the
prediction intervals for the data target according to the MIS, using
a similar format to the RMSE plot in \cref{fig:airline-data-rmse}.
The results shows that AutoGP produces more efficient prediction intervals.
For $T=144$, these reductions reach statistical significance in eight of the
twelve horizons.
For $T=36$, AutoGP significantly improves on X-13 from
$h=4$ through $h=10$, and at $T=72$, the interval comparison is mixed, with one
horizon favoring X-13.
Taken together, these results indicate that the AutoGP model not only
provides more accurate point forecasts of the observed data but also,
across most conditions, delivers improved
prediction intervals vis-a-vis the X-13 model.

\subsubsection{Seasonal Target}
\label{sec:simulation-results-seasonal}

We next evaluate predictions of the seasonal target.
Because the seasonal
amplitude $r(t) $grows over time while the innovation variance remains
fixed, the seasonal signal becomes progressively stronger relative to noise
as the training length increases.
Thus, $T=36$, $72$, and $144$ provide increasingly informative settings for
recovering the seasonal component.

\begin{figure}[!h]
\centering
\includegraphics[width=\textwidth]{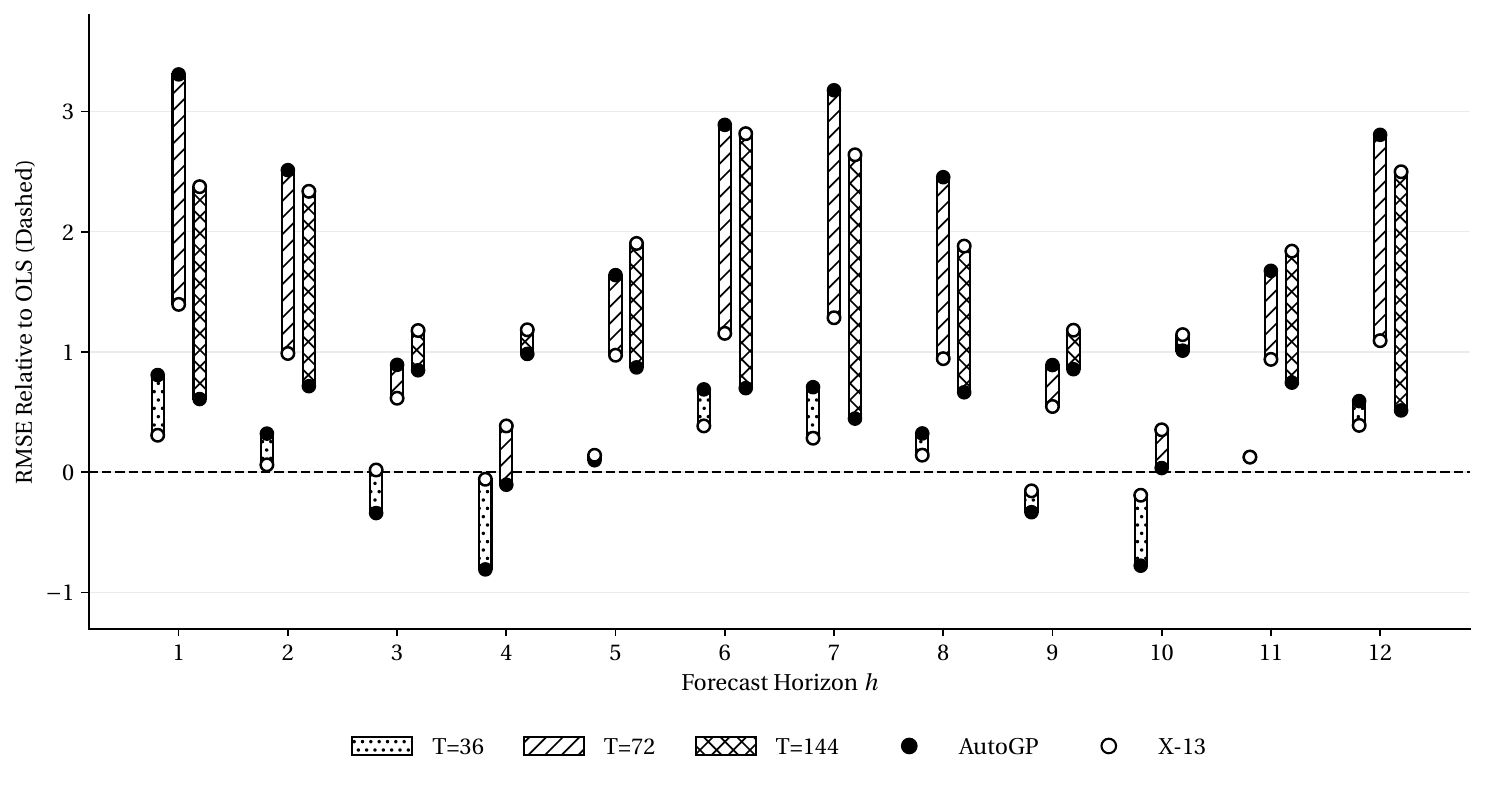}
\caption{Relative RMSE performance for the seasonal target.}
\label{fig:airline-seas-rmse}
\end{figure}

In \cref{fig:airline-seas-rmse}, the RMSE values for predicting the
seasonal target highlight a tradeoff between flexibility and imposed
seasonality.
For the short sample, $T=36$, the seasonal-target results are mixed.
AutoGP significantly improves on X-13 at $h\in\set{3,4,9,10}$, while X-13
significantly improves on AutoGP at six horizons.
In this setting, the seasonal signal is weakest and so neither approach
uniformly dominates.

For each of the $N=200$ datasets, we compute the posterior probability
$\Pr\!\left(m_{\mathrm{per}}\neq 0\mid Y(\bt)=y(\bt)\right)$
that seasonality is present via \cref{eq:query-prob-seasonality}.
For $T=36$, AutoGP assigns negligible posterior mass to
periodic structure, as the seasonal signal is weak relative to the
noise.
Conversely, as X-13 is designed to extract seasonality from data,
its behavior is equivalent to a dogmatic prior heavily weighting the existence
of seasonality in a probabilistic modeling sense.

At the intermediate sample size, $T=72$, X-13 attains lower seasonal error than
AutoGP at most horizons, with statistically significant reductions in ten of
the twelve horizons.
AutoGP significantly improves on X-13 only at $h\in\{4,10\}$.
AutoGP detects seasonality in roughly 81\% of replications, but at this
intermediate training length it tends to distribute the periodic variation
across competing kernel structures rather than the periodic kernel.
This flexibility yields a less-accurate estimate of the seasonal component,
despite more accurately forecasting the data target at $T=72$.
This gap illustrates the central tradeoff of our approach.
X-13 imposes a more fixed seasonal structure by construction, which enables it
to identify a moderate seasonal signal relative to AutoGP.

At the long sample, $T=144$, the ranking reverses.
AutoGP significantly reduces seasonal squared error relative to
X-13 in ten of the twelve horizons, with no
horizon significantly favoring X-13.
Once the sample is sufficiently informative, the flexibility of
probabilistic model discovery outweighs the advantage of imposing
seasonality ex ante, yielding statistically significant RMSE reductions.

\begin{figure}[!h]
\includegraphics[width=\textwidth]{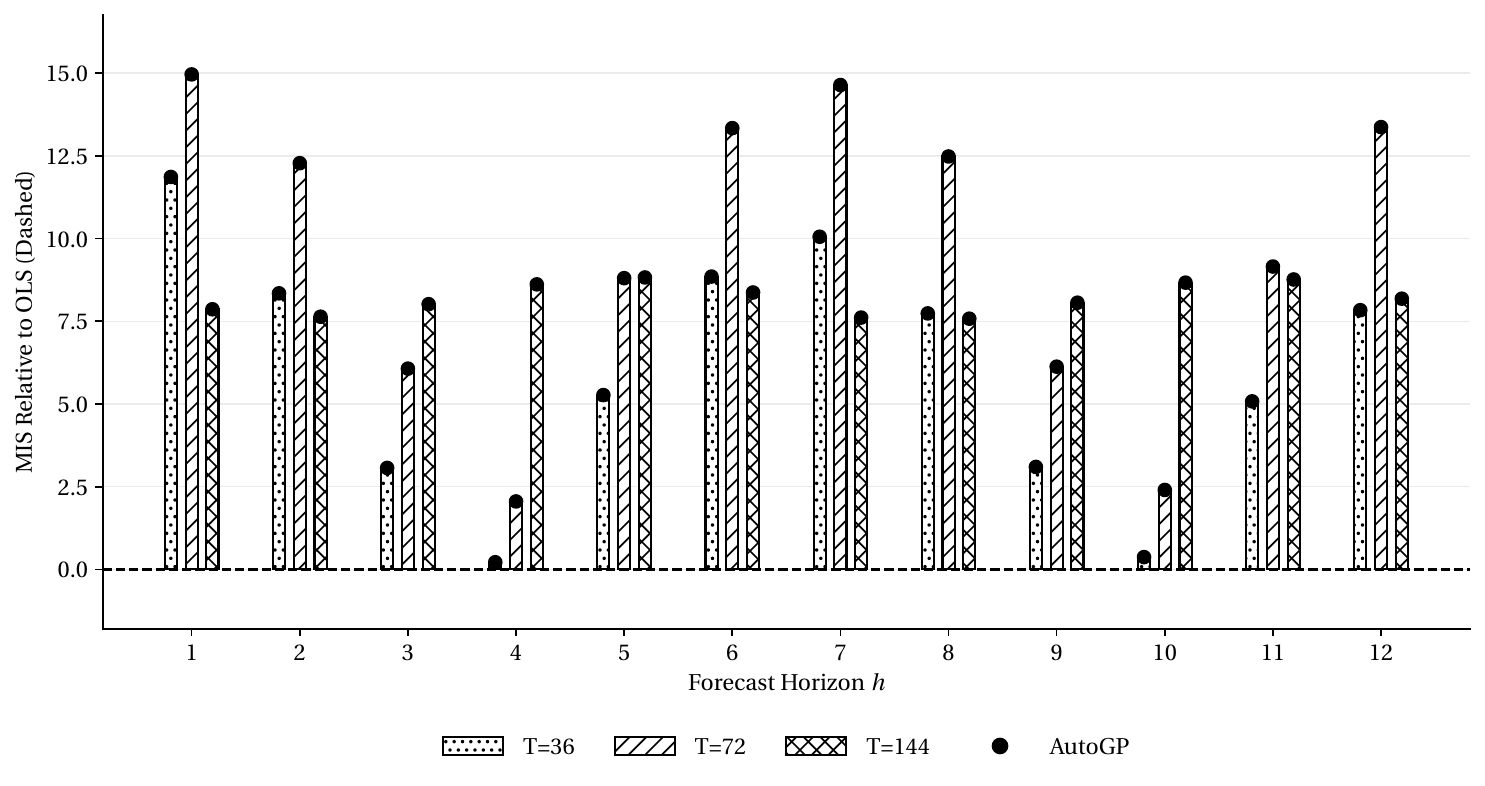}
\caption{Relative MIS performance for the seasonal target.
Note that X-13 is excluded from this comparison due to the lack of valid
predictive intervals for its seasonal factors.
}
\label{fig:airline-seas-mis}
\end{figure}

\Cref{fig:airline-seas-mis} provides the interval-forecast analogue
for the seasonal component using the MIS.
Here the comparison is necessarily limited to AutoGP because X-13 does
not provide usable predictive intervals for the seasonal factor in
this exercise, and so the latter is excluded from \cref{fig:airline-seas-mis}.
Relative to the OLS gold standard, AutoGP's seasonal MIS remains
substantially larger, although this result is expected given the favorable
nature of OLS, which has access to the exact seasonal structure.
To contrast \cref{fig:airline-data-mis},
where AutoGP has moderate relative MIS for the data target,
with \cref{fig:airline-seas-mis}, where AutoGP has substantially
larger relative MIS for the seasonal target,
it is useful to compare the OLS gold-standard forecast-error variances
and corresponding intervals used in the MIS calculation.
For the normalized (demeaned) seasonal target, redefine
$\mathbf q_{T+h} \leftarrow \mathbf q_{T+h} -\frac{1}{12}\sum_{r=1}^{12}\mathbf q_{T+r}$,
so that
$\widetilde s(T+h)=\mathbf q_{T+h}^\top\beta$
and
$\widehat{\widetilde s}(T+h)=\mathbf q_{T+h}^\top\widehat\beta$.
We have
\begin{alignat}{2}
\Var{\widehat Y(T+h)- Y(T+h)}
\quad&= \mathbf z_{T+h}^{\top}\Var{\widehat\beta-\beta}\mathbf z_{T+h} +\sigma_\varepsilon^2
\quad&&=\sigma_\varepsilon^2\left(1+\mathbf z_{T+h}^{\top}(Z^\top Z)^{-1}\mathbf z_{T+h}\right),
\\
\Var{\widehat{\widetilde{s}}(T+h)- \widetilde{s}(T+h)}
\quad&= \mathbf q_{T+h}^{\top}\Var{\widehat\beta-\beta}\mathbf q_{T+h}
\quad&&= \sigma_\varepsilon^2\mathbf q_{T+h}^{\top}(Z^\top Z)^{-1}\mathbf q_{T+h},
\end{alignat}
where
$Z \defas \left[ \mathbf{z}_1 \; \dots \; \mathbf{z}_T \right]^\top$
is the OLS design matrix.
The first variance is over both the in-sample innovations
$\varepsilon(\bt)$, which determine
$\widehat\beta$, and the held-out innovation $\varepsilon(T+h)$ at horizon $h$.
The second variance is only over the in-sample innovations through
$\widehat\beta$. The corresponding OLS 95\% prediction intervals are
\begin{align}
\widehat Y(T+h)\pm t_{T-6,0.975}\sqrt{\widehat V_{Y,h}},
&&
\widehat V_{Y,h} &\defas \widehat\sigma_\varepsilon^2\left(1+\mathbf z_{T+h}^{\top}(Z^\top Z)^{-1}\mathbf z_{T+h}\right),
\\
\widehat{\widetilde{s}}(T+h)\pm t_{T-6,0.975}\sqrt{\widehat V_{s,h}},
&&
\widehat V_{s,h} &\defas \widehat\sigma_\varepsilon^2\mathbf q_{T+h}^{\top}(Z^\top Z)^{-1}\mathbf q_{T+h}.
\end{align}
Thus the data-target interval must account for the held-out innovation,
whereas the seasonal-target interval reflects only parameter
uncertainty, which gives a much sharper reference point.
Because the relative MIS calculation normalizes by the OLS MIS separately for
each target, AutoGP's relative MIS can appear much larger for the seasonal
target partly because the OLS seasonal denominator is significantly
smaller.


\section{Empirical Analysis}
\label{sec:empirical}

We now evaluate how our seasonal adjustment method performs on real-world
economic data.
Our analysis examines eight monthly U.S. macroeconomic series
selected to span labor markets, consumer prices, industrial production,
retail spending, residential construction, manufacturing demand, and
business inventories, as shown in \cref{table:series}.
Together, the series provide a heterogeneous panel that differs substantially
in persistence, volatility, seasonal strength, and susceptibility to large
economic shocks.

\begin{table}[!h]
\centering
\setlength{\tabcolsep}{6pt}
\begin{threeparttable}
\caption{Monthly macroeconomic series.}
\label{table:series}
\footnotesize
\begin{tabularx}{\textwidth}{l>{\ttfamily}l>{\ttfamily}llXc}
\toprule
\textbf{Series}
& \normalfont{\textbf{NSA Series}}
& \normalfont{\textbf{SA Counterpart}}
& \textbf{Source}
& \textbf{Units}
& \textbf{Analysis Sample}
\\
\midrule
Apparel CPI
& CUUR0000SAA2
& CUSR0000SAA2
& BLS
& Index (1982--84 = 100)
& 1990:01--2025:09
\\
\addlinespace
Business Inventories
& TOTBUSIMNSA
& BUSINV
& Census
& Dollars (Millions)
& 1992:01--2025:09
\\
\addlinespace
Core CPI
& CPILFENS
& CPILFESL
& BLS
& Index (1982--84 = 100)
& 1990:01--2025:09
\\
\addlinespace
Durable-Goods Orders
& UMDMNO
& DGORDER
& Census
& Dollars (Millions)
& 1992:02--2025:09
\\
\addlinespace
Housing Starts
& HOUSTNSA
& HOUST
& Census/HUD
& Units (Thousands)
& 1990:01--2025:09
\\
\addlinespace
Industrial Production
& IPB50001N
& INDPRO
& Federal Reserve G.17
& Index (2017 = 100)
& 1990:01--2025:09
\\
\addlinespace
Payroll Employment
& PAYNSA
& PAYEMS
& BLS
& Persons (Thousands)
& 1990:01--2025:09
\\
\addlinespace
Retail Sales
& RSAFSNA
& RSAFS
& Census
& Dollars (Millions)
& 1992:01--2025:09
\\
\bottomrule
\end{tabularx}
\begin{tablenotes}[para,flushleft]
\footnotesize
\item[\emph{Notes:}]
The listed series are the not-seasonally-adjusted (NSA) monthly series
used as model inputs.
The corresponding official seasonally adjusted (SA) series are listed
for comparison but are not used to estimate the forecast models.
\end{tablenotes}
\end{threeparttable}
\end{table}

These series include several of the most closely watched measures of U.S.
economic activity.
Payroll Employment and Core CPI summarize labor-market conditions and
underlying consumer-price inflation.
Retail Sales and Durable-Goods Orders provide information about household
spending and manufacturing demand.
Housing Starts and Industrial Production are important indicators of
cyclical activity.
The panel also spans substantial differences in seasonal behavior (see
\cref{appx:data}).
Seasonality is a prominent feature of Retail Sales, Housing Starts,
Payroll Employment, and Durable-Goods Orders, while
Core CPI and Business Inventories are smoother and more trend dominated.
This variation allows us to evaluate whether the performance of Seasonal
AutoGP relative to X-13 depends on the strength or shape of the seasonal
component.\footnote{We separately examine the relationship between
seasonal adjustment and official data revisions using release-aligned
ALFRED vintages.
Applying X-13 to the NSA history available at each release closely reproduces
official SA revisions for Industrial Production, Business Inventories,
Payroll Employment, Retail Sales, and Durable-Goods Orders, but less so for Core CPI,
Apparel CPI, and Housing Starts.}

\subsection{Evaluation Setup}
\label{sec:empirical-setup}

For each time series $i$, let $y_i^{\rm NSA}(t)$ denote the
not-seasonally-adjusted (NSA) observation in levels in month $t$, and define
$y_i(t) \defas \log y_i^{\rm NSA}(t)$.
The available dates are shown in the Analysis Sample column of
\cref{table:series}.
For each time series $i$, we define the forecast origins to be the December
months 2000:12, 2001:12, $\dots$, 2023:12.
For every origin $v$, each method is estimated using data available through
$t=1,\dots,v$ and produces a twelve-month forecast $y_i(v + h)$ for
$h=1,\dots,12$, i.e., January to December of the following year.
\Cref{appx:preproc} describes standard series-specific transformations (e.g.,
removing calendar-day effects and applying first differences where necessary) that ensure a direct
comparison between AutoGP and X-13, and the logarithmic scale on which
forecast losses are computed.\footnote{We use X-13ARIMA-SEATS Version 1.1 Build
62.
At each origin, X-13 automatically selects the transformation and ARIMA
model, applies AIC tests for trading-day and Easter regressors, and searches
for additive outliers, level shifts, and temporary changes.
The ARIMA residual diagnostic uses a maximum lag of $36$, the forecast
module produces twelve leads with $95$ percent intervals, and the X-11
seasonal moving average is selected using \texttt{seasonalma=msr}.}

\subsection{Forecast Results}
\label{sec:empirical-forecast}

\Cref{table:panel-best-autogp-x13} reports, for each series, the AutoGP and
X-13 RMSE and MIS, together with the AutoGP-to-X-13 ratio of each
(ratios below one favor AutoGP).
For example, Housing Starts has an AutoGP RMSE of $0.124$ versus
$0.157$ for X-13 (a ratio of $0.789$) and an AutoGP MIS of $0.778$
versus $0.797$ (a ratio of $0.976$).
A log RMSE of $0.124$ for Housing Starts corresponds approximately to a
relative forecast error of $12.4$ percent, versus $15.7$ percent for X-13.
For illustration, at a level of $1.4$ million starts at a seasonally
adjusted annual rate, this difference corresponds to roughly $46{,}000$
starts on the annualized scale.
For Payroll Employment, an AutoGP RMSE of $0.021$ on a level near
$155$~million corresponds to a typical error of about $3.3$~million
persons.


\begin{table}[!h]
\centering
\begin{threeparttable}
\caption{AutoGP versus robust automated X-13 forecast distributions.}
\label{table:panel-best-autogp-x13}
\small
\begin{tabular}{lrrrrrr}
\toprule
Series
& AutoGP RMSE
& X-13 RMSE
& RMSE ratio
& AutoGP MIS
& X-13 MIS
& MIS ratio
\\
\midrule
Apparel CPI
& 0.026 & 0.025 & 1.013
& 0.136 & 0.126 & 1.083
\\
Business Inventories
& 0.029 & 0.031 & 0.929
& 0.182 & 0.201 & 0.907
\\
Core CPI
& 0.006 & 0.007 & 0.847
& 0.044 & 0.066 & 0.673
\\
Durable-Goods Orders
& 0.075 & 0.077 & 0.977
& 0.520 & 0.556 & 0.935
\\
Housing Starts
& 0.124 & 0.157 & 0.789
& 0.778 & 0.797 & 0.976
\\
Industrial Production
& 0.030 & 0.032 & 0.963
& 0.212 & 0.213 & 0.996
\\
Payroll Employment
& 0.021 & 0.022 & 0.962
& 0.162 & 0.180 & 0.899
\\
Retail Sales
& 0.037 & 0.037 & 1.004
& 0.241 & 0.315 & 0.763
\\
\bottomrule
\end{tabular}
\begin{tablenotes}[para]
\footnotesize
\item[\emph{Notes:}]
The table reports RMSE and the mean interval score (MIS) for $95$ percent
prediction intervals, in log units, pooled over horizons $h=1,\ldots,12$
and the $23$ complete forecast origins. Ratios are AutoGP divided by X-13, so
values below one favor AutoGP. Four predictions for X-13 Housing Starts
are nonpositive at 2008:12 and therefore cannot be transformed into log interval, so
all measurements for this incomplete forecast origin are dropped.
\end{tablenotes}
\end{threeparttable}
\end{table}

\Cref{table:panel-best-autogp-x13} shows that AutoGP has lower RMSE for
six of the eight series.
The exceptions are Retail Sales, where the ratio is $1.004$ (essentially a
tie), and Apparel CPI, where X-13 is modestly better (ratio $1.013$).
Similarly, AutoGP has lower MIS for seven of the eight series.
The largest MIS improvements are for Core CPI (ratio $0.673$,
approximately $33$ percent reduction) and Retail Sales (ratio $0.763$),
followed by Payroll Employment ($0.899$) and Business Inventories
($0.907$).
Apparel CPI is the only series for which X-13 performs better on both
criteria, although the differences remain moderate.
Our small evaluation sample means that most individual improvements are not
statistically distinguishable from zero.
This effect may arise from limited statistical power rather than an absence
of forecast improvement, as the loss ratios fall at or below one for nearly
every series.

\subsection{Residual Seasonality}
\label{sec:empirical-freq}

We next evaluate how completely X-13 and AutoGP remove conventional
fixed-frequency seasonal variation from the analyzed series.
\Citet{Findley:2017} observe that a fundamental deficiency in seasonal
adjustment is detectable seasonality after adjustment, and define various
diagnostics that can be used to quantify the magnitude of residual
seasonality.
These diagnostics are useful for characterizing the decomposition, but they
are not forecast loss functions and should not be interpreted as the
primary measure of empirical performance.
A procedure can, in principle, remove nearly all fixed-frequency seasonal
variation without improving its prediction of future observations.
Conversely, a model can leave detectable periodic variation while providing
accurate and well-calibrated forecasts of the observed series.%
\footnote{Moreover, these residual-seasonality diagnostics should themselves be
interpreted cautiously.
For the series analyzed in log levels, the trend must also be estimated
and removed before power at the seasonal frequencies can be measured.
Imperfect detrending and leakage from large low-frequency movements can
either obscure genuine seasonal variation or generate apparent power
near the seasonal harmonics.
Consequently, failure to eliminate all measured seasonal-frequency
variation need not imply that the predictive model has failed to learn
the economically relevant seasonal structure.}

For each series $i$ and origin $v$,
let $P_{i,v}^{\rm SA}$ denote the absolute spectral power of the
seasonally adjusted series in frequency bands surrounding the six monthly
seasonal harmonics (see \cref{appx:frequency}),
and define
$P_{i,v}^{\rm NSA}$ analogously for the not-seasonally-adjusted series.
The residual seasonal power is defined as the ratio
$P_{i,v}^{\rm SA} / {P_{i,v}^{\mathrm{NSA}}}$.
We calculate this statistic across all six monthly seasonal harmonics
and separately using only the annual fundamental, for both X13 and AutoGP.
The all-harmonic statistic is the more comprehensive diagnostic because a
monthly seasonal pattern need not be sinusoidal and can therefore place
substantial power at higher seasonal harmonics.
The annual-fundamental statistic computes the contribution at the
$12$-month fundamental only.
Following \citet{Findley:2017},
we provide a complementary time-domain diagnostic by regressing the transformed
adjusted series on the identifiable sine and cosine terms associated
with the six monthly seasonal harmonics, and report the resulting
$R^2$.
Low residual power and residual harmonic $R^2$ indicate that the
adjustment has removed seasonal-frequency variation.
For Core CPI and Business Inventories, the diagnostics are computed
from monthly log changes, matching the transformation used in
estimation.
Linearly detrended log levels are used for the remaining series.

\begin{table}[!h]
\centering
\begin{threeparttable}
\caption{Residual fixed-frequency seasonal variation after seasonal adjustment.}
\label{tab:seasonal-removal}
\small
\setlength{\tabcolsep}{4.5pt}
\begin{tabular}{lrrrrrr}
\toprule
&
\multicolumn{2}{c}{\begin{tabular}{c}Residual Power Ratio\\(All Harmonics)\end{tabular}}
&
\multicolumn{2}{c}{\begin{tabular}{c}Residual Power Ratio\\(Annual Fundamental)\end{tabular}}
&
\multicolumn{2}{c}{\begin{tabular}{c}Residual Harmonic $R^2$ \\ (Time Domain)\end{tabular}}
\\
\cmidrule(lr){2-3}
\cmidrule(lr){4-5}
\cmidrule(lr){6-7}
Series
& AutoGP & X-13
& AutoGP & X-13
& AutoGP & X-13
\\
\midrule
Apparel CPI
& 0.3\% & 0.7\%
& 2.7\% & 11.3\%
& 0.6 & 1.4
\\
Business Inventories
& 1.4\% & 0.9\%
& 6.1\% & 5.2\%
& 1.7 & 1.5
\\
Core CPI
& 3.9\% & 2.6\%
& 4.3\% & 3.0\%
& 3.3 & 3.5
\\
Durable-Goods Orders
& 7.4\%  & 4.6\%
& 61.0\% & 39.0\%
& 0.8 & 1.6
\\
Housing Starts
& 4.5\% & 2.4\%
& 1.6\% & 1.0\%
& 2.2 & 2.2
\\
Industrial Production
& 5.2\%  & 3.8\%
& 15.2\% & 9.3\%
& 0.7 & 0.8
\\
Payroll Employment
& 0.5\% & 0.3\%
& 0.9\% & 0.5\%
& 0.9 & 1.1
\\
Retail Sales
& 1.0\% & 0.3\%
& 4.0\% & 2.0\%
& 1.6 & 1.4
\\
\bottomrule
\end{tabular}
\begin{tablenotes}[para]
\footnotesize
\item[\emph{Notes:}]
Entries are percentages and are medians across the $25$ December
estimation origins from 2000:12 through 2024:12.
For Core CPI and Business Inventories, the statistics are computed from
monthly log changes, while for the remaining series, the statistics are
computed from linearly detrended log levels.
``All Harmonics'' is
the ratio of the absolute spectral power (in bands surrounding the six monthly seasonal harmonics)
in the seasonally adjusted data and in the not-seasonally-adjusted data.
``Annual Fundamental'' is defined analogously using only the band surrounding
the 12-month annual fundamental.
``Residual Harmonic $R^2$'' is the fraction of variation in the transformed
adjusted series explained by the identifiable sine and cosine terms at the
six monthly seasonal harmonics.
\end{tablenotes}
\end{threeparttable}
\end{table}

Both AutoGP and X-13 remove nearly all conventional fixed-frequency
seasonal variation across the panel under both the frequency-domain and
time-domain diagnostics (\cref{tab:seasonal-removal}).
The sole exceptions are Industrial Production and Durable-Goods Orders at
the annual fundamental, which is not a dominant frequency for these series
(\cref{appx:frequency}).
The differences in residual power across all six harmonics are also small.
Seasonal AutoGP produces lower residual power than X-13 for Apparel
CPI, while X-13---which is designed to remove specific harmonics---is
lower for the other seven series.
The residual harmonic $R^2$ is consistent with these results, as
the maximum difference is at most $0.8$ percentage points across all of the series.
These results indicate the forecast gains in
\cref{table:panel-best-autogp-x13} using \SGP{} coexist with largely
comparable removal of conventional seasonal effects as X-13.

\subsection{Revision Timing \& Risk}
\label{sec:empirical-revision}
An advantage of Seasonal AutoGP relative to other approaches
is the ability to quantify uncertainty about the seasonal adjustment
in a coherent manner.
Our Revision Risk query of \cref{sec:seasonality-queries} addresses
an issue that is particularly problematic for X-13.
The X-13 moving-average filter (dubbed X-11) estimates the seasonal pattern for
a calendar month as a weighted average of nearby same-month observations.
Any extreme observation, therefore, is partially absorbed into the seasonal factors
assigned to that month in adjacent years.
Substantial outliers can impact seasonal adjustments for up to five years
\citep{BLSSeasonalAdjustment:2026}.
We first document realized revisions to the X-13 seasonal adjustment over the pandemic period,
demonstrating the challenges with communicating seasonality in real time using this
methodology.
We then show how Seasonal AutoGP offers a probabilistic assessment of such
quantities.

\subsubsection{X-13 Seasonal Revisions}
\label{sec:empirical-revision-x13}
Consider
the fixed current-vintage history
$y^{\rm NSA}_i(1),\ldots,y^{\rm NSA}_i(T_i)$
of NSA levels through 2025:09 for each series $i$.
Let $c_i$ denote the corresponding COVID ``shock month'', which is
May 2020 for Business Inventories and April 2020 for the remaining series.
We use the rolling observations
$y^{\rm NSA}_i(1),\ldots,y^{\rm NSA}_i(e)$ to build
an analysis run with expanding endpoints $e$.
In particular, when moving from $e$ to $e+1$, we hold earlier observations
fixed, add one month, and re-estimate X-13, including its automatically
selected transformation, regression model, outliers, and filters.
X-13 identifies outliers by iteratively adding additive-outlier,
level-shift, and temporary-change regressors with $t$-statistics exceeded a
specified critical value that grows with sample size \citep{X13RefManual:2024}, and
adjusts the flagged observations before the seasonal filter is applied.
As these quantities change with each new observation,
so does the seasonal component assigned to the shock month.

As X-13 and \SGP{} admit different seasonal decompositions,
it is essential to quantify their revisions on the same scale.
For the \SGP{} decomposition, we model the log NSA data
$\log\left(y_i^{\rm NSA}(t)\right) = y_i(t) = s_i(t) + u_i(t) + \epsilon_i(t)$ following \cref{eq:autogp-xsu-decomposition},
so the extracted seasonal component $s_i(t)$ is already measured in log units.
Equivalently, $s_i(t)$ from \SGP{} is the log ratio of the NSA level to the
corresponding \SGP-adjusted level.
To place the seasonal component from X-13 on the same log scale,
we let $F_{i}^{(e)}\!(t)$ be the final one-centered
seasonal component in a multiplicative decomposition
and $C_{i}^{(e)}\!(t)$ the seasonal component
in an additive decomposition
(i.e., \texttt{d10}, as described in
\citet[Section 7.19, Table 7.51]{X13RefManual:2024})
and define the X-13 log seasonal component as
\begin{equation}
s_i^{(e)}(t)
\defas
\begin{cases}
\log F_{i}^{(e)}\!(t)
&
\text{multiplicative X-13 decomposition}
\\[4pt]
\log\left(\dfrac{\mathrm{SA}_{i}^{(e)}\!(t) + C_{i}^{(e)}\!(t)}{\mathrm{SA}_{i}^{(e)}\!(t)} \right)
&
\text{additive X-13 decomposition},
\end{cases}
\label{eq:x13-seasonal-effect}
\end{equation}
where $\mathrm{SA}_{i}^{(e)}\!(t)$
is the corresponding X-11 seasonally adjusted level (i.e., \texttt{d11}).
Holding the shock month $c_i$ of series $i$ fixed,
we report the absolute revision each endpoint
$e = c_i + h$ in log percentage points:
\begin{align}
Q_i(h)\defas 100 \abs*{s_{i}^{(c_i+h)}(c_i)-s_{i}^{(c_i)}(c_i)},
\label{eq:x13-covid-seasonal-revision-path}
\end{align}
which for small revisions is approximately equal to the absolute percentage
revision of the seasonal component.
In order to contextualize these values, we also compute
the historical revision magnitudes by
constructing,
for each pre-COVID shock month
$t=2012{:}01,\ldots,2019{:}12$,
the analogous $Q_{i,t}^{\mathrm{pre}}(h)$ whenever $t+h\leq2019{:}12$.
%

\begin{figure}[p]
\centering
\includegraphics[width=\textwidth]{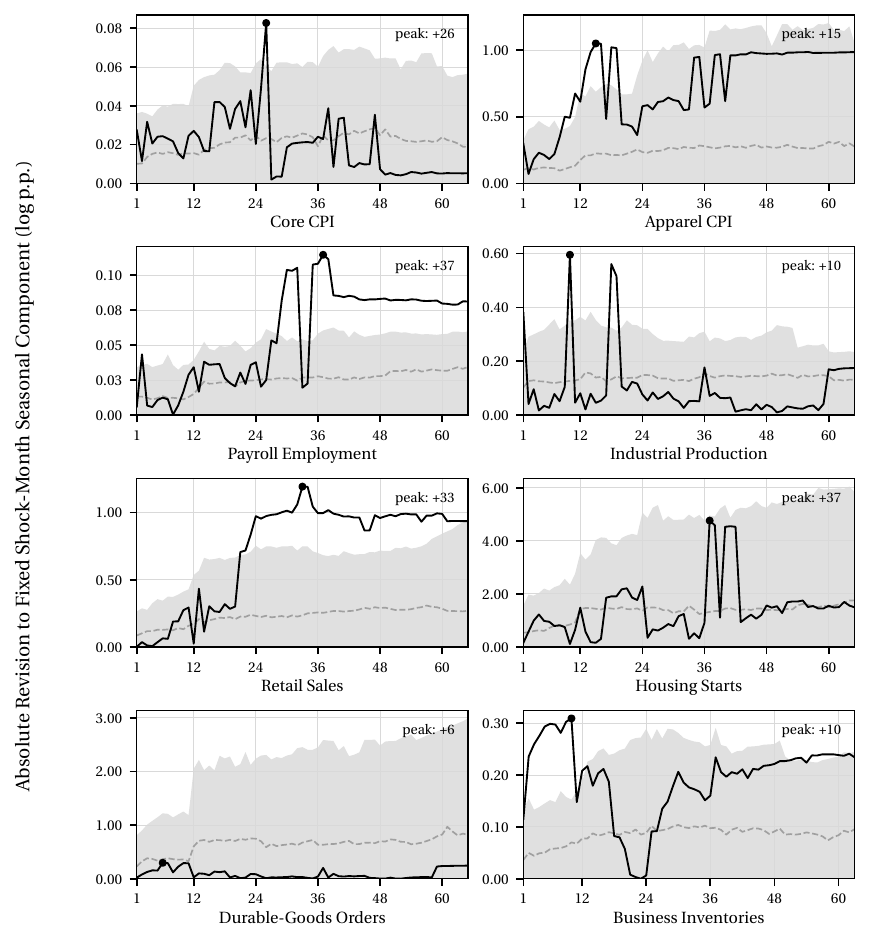}
\caption{X-13 revision paths for the seasonal component assigned to the
COVID shock month. The solid line plots the absolute
revision after $h$ additional monthly observations
in log percentage points. The dashed line is the
median of the corresponding pre-COVID revisions at the same horizon, and
the gray region spans their 5th--95th percentiles. Peak labels report the
horizon at which the COVID revision is largest.}
\label{fig:x13-revisions}
\end{figure}

\Cref{fig:x13-revisions} plots the median pre-COVID seasonal revision
with the shaded gray region accounting for the 5th--95th percentile of
pre-COVID revisions.
The black line plots $Q_i(h)$, which measures how much X-13 has revised the
seasonal component initially assigned to the shock observation after
$h$ additional releases, on the log percentage points scale.
For all but Durable Goods, peak revisions occur at least 10 months after
the initial shock.
Peak revisions generally lie above the 95th percentile of pre-COVID
revisions, with Housing Starts approximately at that percentile.
For Payroll Employment, Retail Sales, and Housing Starts, the peak revision
occurs more than two years after the shock.
Not surprisingly, X-13 categorizes many observations as outliers during and
slightly after the pandemic.
For Payroll Employment, Industrial Production, and Retail Sales, X-13
initially attributes much of the COVID shock to outliers.
These fitted outlier effects are approximately $15$--$20$ percent in
absolute magnitude, while the seasonal factor assigned to the shock month
differs by less than approximately $0.1$ percent from that assigned to the
same calendar month one year earlier.
This initial allocation of observations to outliers helps explain why
large X-13 revisions to the seasonal component emerge only after many
additional observations arrive.

\subsubsection{Seasonal AutoGP Revision Risk}
\label{sec:empirical-revision-autogp}

Rather than quantify revisions retrospectively as in X-13,
the Revision Risk of \cref{sec:seasonality-queries} takes advantage
of the Bayesian nature of \SGP{} to provide an estimate of how
much a seasonal adjustment reported today may change in the future.
At each monthly origin $e=c_i,\ldots,c_i+12$, we draw $B=1{,}000$ possible
paths $y_{i,e+1:e+6}^{(b)}$ ($b = 1, \dots, B$) for the next six
observations from the current posterior predictive distribution.
For each path, we then evaluate the six-month revision defined in \cref{eq:defn-revision},
using a fixed 24-month normalization window ending at the shock month $c_i$.
We write $R^{(b)}_{i,e}$ for the resulting draws of the revision at the
shock month associated with origin $e$.
The random variables $R_{i,e}^{(1:B)}$ are thus i.i.d.~draws from the
Revision Risk distribution~\cref{eq:revision-risk-distribution}
available at origin $e$, before the next six
observations are known.

Our first experiment addresses the following question:
\textit{In real time, how much did seasonal uncertainty increase as COVID unfolded?}
For each origin $e$, let
$
W_{i,e}^{90} \defas
  100\left(
    \operatorname{Quantile}_{0.95}\!\left(R^{1:B}_{i,e}\right)-\operatorname{Quantile}_{0.05}\!\left(R^{1:B}_{i,e}\right)
    \right)
$
denote the width of its $90$-percent interval.
In order to normalize the magnitudes of revision risks and compare them across series,
we define the post-shock \emph{Revision Risk Index}
for series $i$ at the post-COVID origin $e$ as
\begin{align}
\operatorname{RRI}_{i,e}
\defas
\frac{W_{i,e}^{90}}{\underset{u \in \set{2000:12,\dots,2019:12}}{\operatorname{median}}W_{i,u,e-c_i}^{90,\mathrm{pre}}}
&& (c_i \le e \le c_i + 12).
\label{eq:autogp-revision-risk-index}
\end{align}
The term $W_{i,u,k}^{90,\mathrm{pre}}$ denotes the corresponding
$90$-percent revision-interval width at historical December origin $u$
for a target $k$ months old.
Thus, setting $k=e-c_i$, the denominator in
\cref{eq:autogp-revision-risk-index} is the median pre-COVID revision risk
for the same series and target age.
The Revision Risk Index is dimensionless.
A value of one gives the historical median, while values above the
corresponding historical $95$th percentile identify unusually high Revision
Risk.

\begin{table}[!t]
\centering
\small
\renewcommand{\arraystretch}{1.08}
\begin{threeparttable}
\caption{Seasonal AutoGP Revision Risk Index during the COVID pandemic.}
\label{table:autogp-revision-risk}
\begin{tabular}{lrrrrrr}
\toprule
Series & Shock & $+3$ & $+6$ & $+12$ & Peak (origin) & Above 95th \\
\midrule
Core CPI & $\mathbf{3.8}$ & $\mathbf{5.5}$ & $\mathbf{2.3}$ & $\mathbf{1.9}$ & $\mathbf{5.9}\ (+2)$ & $12/13$ \\
Apparel CPI & $\mathbf{1.6}$ & $\mathbf{4.2}$ & $\mathbf{1.5}$ & $1.3$ & $\mathbf{4.2}\ (+3)$ & $9/13$ \\
Payroll Employment & $\mathbf{5.6}$ & $\mathbf{329.2}$ & $\mathbf{168.7}$ & $\mathbf{3.8}$ & $\mathbf{615.0}\ (+2)$ & $13/13$ \\
Industrial Production & $\mathbf{1.8}$ & $\mathbf{6.3}$ & $\mathbf{8.4}$ & $\mathbf{1.6}$ & $\mathbf{20.2}\ (+10)$ & $13/13$ \\
Retail Sales & $1.3$ & $\mathbf{44.9}$ & $\mathbf{4.2}$ & $\mathbf{1.7}$ & $\mathbf{44.9}\ (+3)$ & $12/13$ \\
Housing Starts & $0.7$ & $\mathbf{55.0}$ & $\mathbf{2.1}$ & $\mathbf{2.2}$ & $\mathbf{55.0}\ (+3)$ & $9/13$ \\
Durable-Goods Orders & $0.9$ & $\mathbf{5.3}$ & $\mathbf{1.9}$ & $\mathbf{2.0}$ & $\mathbf{7.0}\ (+4)$ & $9/13$ \\
Business Inventories & $1.1$ & $\mathbf{10.0}$ & $2.0$ & $1.1$ & $\mathbf{13.8}\ (+4)$ & $5/13$ \\
\bottomrule
\end{tabular}
\begin{tablenotes}[para]
\footnotesize
\item[\emph{Notes:}]
An index of one is the pre-COVID median for the same series and target age.
Bold entries exceed the corresponding historical 95th
percentile. Parentheses give the origin of the peak in post-shock months.
The final column counts exceedances across the 13 post-COVID endpoints.
\end{tablenotes}
\end{threeparttable}
\end{table}

\Cref{table:autogp-revision-risk} shows that the Revision Risk rose
substantially as the pandemic unfolded.
At the initial shock, four of the eight series already exceeded their
historical $95$th percentile (bold entries).
Three months later, the median Revision Risk was $8.1$ and all eight series
exceeded their historical $95$th percentile.
Revision Risk remains elevated thereafter with seven series exceeding the
threshold at $+6$ and six still exceeding it at $+12$, when the median
indices are $2.2$ and $1.8$, respectively.
Overall, $82$ of the $104$ COVID observations lie above the corresponding
pre-COVID $95$th percentile, and every series does so at least once.
All eight series attain their maximum after the shock origin
(seven by $+4$, while Industrial Production peaks latest at $+10$).
For Payroll Employment, at $+2$ months, its COVID interval width is $5.766$
log percentage points, compared with a pre-COVID median of
$0.0094$, yielding an index of approximately $615$.
This wide ratio reflects both COVID uncertainty interval and
(more importantly) the narrow pre-COVID benchmark.
\Cref{table:autogp-revision-risk} demonstrates a benefit of
reporting the Revision Risk statistic over this volatile post-COVID period,
as it provides a coherent measure of uncertainty intrinsic to the seasonal adjustment itself.

\paragraph{Calibration}

We next assess the extent to which these higher values of Revision Risk
materialized in the actual data.
Our next experiment asks: \textit{Did the realized revisions to the
seasonal component assigned to the shock month appear plausible under the
Revision Risk distribution available at that time?}

For each origin $e=c_i, \dots, c_{i+6}$, we use the next six observations
$y_{i,e+1:e+6}^{\mathrm{obs}}$ that actually occurred to compute the
realized revision $r^{\rm obs}_{i,e}$.
Using the $B=1000$ ex-ante draws of the revision, we can test whether the
realized revision lies inside the central $90$-percent Revision Risk
interval:
\begin{align}
U_{i,e} \defas \frac{1}{B} \sum_{b=1}^{B} \mathbf{1}\!\set*{ R_{i,e}^{(b)} \leq r_{i,e}^{\mathrm{obs}} } \in [0.05, 0.95]
\label{eq:autogp-realized-revision-predictive-rank}
&& (c_i \le e \le c_i+6).
\end{align}
For example, a value of $U_{i,e}=0.50$ implies that the realized revision was
near AutoGP's predictive median.

\begin{figure}[p]
\centering
\includegraphics[width=\textwidth]{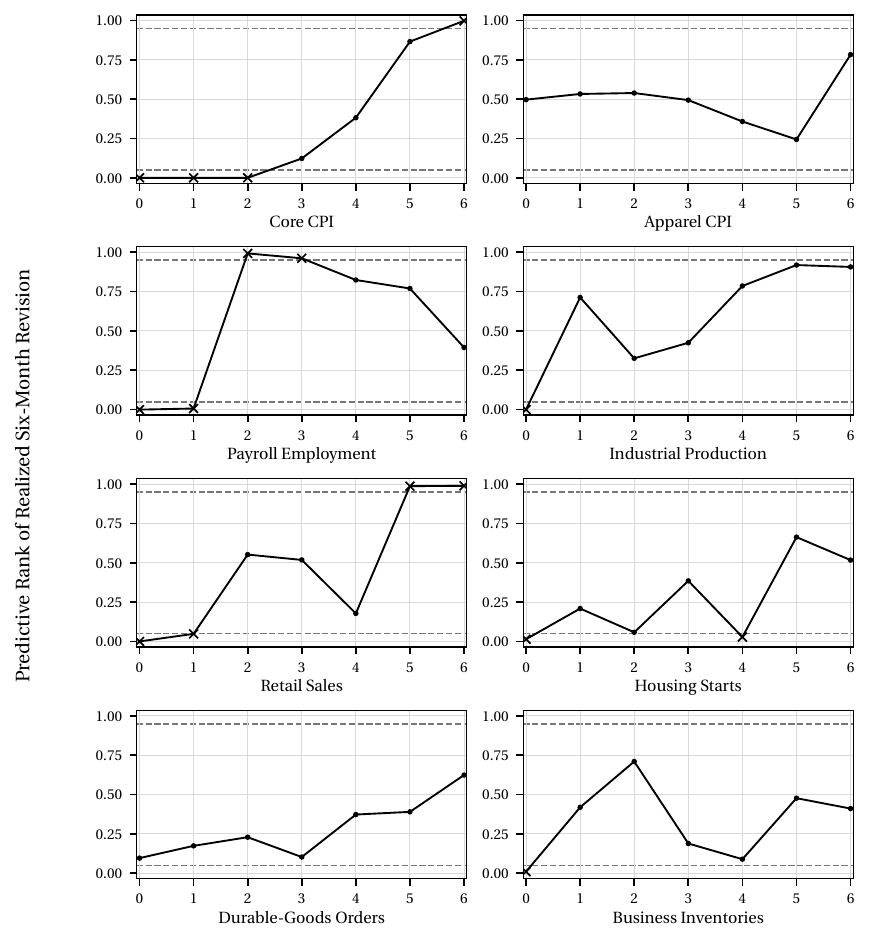}
\caption{Rolling AutoGP Revision Risk diagnostic. At each of the seven
post-COVID origins (+0 to +6 months) on the x-axes, the y-values show the
predictive rank of the revision to the shock month induced by the actual
next six observations within $1{,}000$ draws from the ex-ante distribution.
Crosses mark realized revisions outside the 90 percent central region
(dashed lines).}
\label{fig:autogp-revisions}
\end{figure}

\Cref{fig:autogp-revisions} shows the values of $U_{i,e}$.
At the shock origin, only two of the eight realized revisions fell inside
the central $90$-percent interval, which suggests the initial
predictive distributions did not anticipate the enormous adjustments
that would occur over the next six months.
However, the Revision Risk diagnostic rapidly increased as the pandemic
unfolded.
Five realized revisions fell inside the central $90$-percent interval one
month later, six after two months, and seven after three months.
Seven series are also covered at $+4$ and $+5$, and six at $+6$.
Thus, while the initial observation did not generate a dispersed
enough Revision Risk diagnostic, the distribution adapted as
additional observations arrived.
In total, $40$ of the $56$ rolling realized revisions lie inside their
ex-ante intervals.

Comparing the seasonal revisions of X-13 and Seasonal AutoGP reveals that
the two procedures arrive at similar final seasonal patterns (the median
correlation between the final seasonal components is $0.972$),
but through different incremental revisions.
For \SGP, the largest
observed seasonal adjustment occurs between months two and five for every series;
for five series (Industrial Production, Retail Sales, Housing Starts,
Durable-Goods Orders, Business Inventories), X-13's largest monthly
adjustment instead occurs between months 10 and 12.
The timing of revisions therefore varies more than the final adjustment
size.
Seasonal AutoGP revisions occur as the pandemic sequence is observed,
while the method reports large uncertainty about the revision in real time.
In contrast, X-13 defers large revisions until much later,
and does not report uncertainty with each release.

\section{Conclusion}
\label{sec:conclusion}

This paper introduces a novel probabilistic model discovery framework
for seasonal adjustment of time series called \SGP.
The method uses Gaussian process models to automatically discover seasonal
and nonseasonal components through a data-driven decomposition.
It is based on model hat can express flexible patterns such as
nonstationary and evolving seasonality, while also delivering full
posterior inferences over latent seasonal structure.
This approach yields improved forecast accuracy and uncertainty estimates
compared to conventional baselines, which is particularly valuable when
seasonal patterns shift due to economic shocks.

In a simulation study on classic economic benchmark of airline passenger
data, we find that \SGP{} produces improved point and interval forecasts
compared to X-13.
Its accuracy in recovering the latent seasonal component exhibits an
informative tradeoff.
When the periodic signal in the data is weak, using a hand-specified
imposed periodic structure as in X-13 performs better than attempting to
automatically discover periodic structure using \SGP.
However, once the samples are sufficiently informative so as to distinguish
periodic structure, \SGP{} achieves significantly lower prediction errors.

Using a case study of eight U.S.\ macroeconomic series, we
find that \SGP{} achieves lower forecast errors for six
series and lower interval forecast errors for seven.
Both X-13 and \SGP{} remove similar amounts of fixed-frequency
seasonal variation, which shows that these forecast improvements do not
occur at the expense of weaker removal of conventional seasonal effects that
are common in many economic time series.

A key contribution of \SGP{} is its notion of ``Revision Risk'', which
delivers an ex-ante distribution over the size of future revisions to a
current seasonal adjustment.
In a study of rolling revisions during the COVID-19 pandemic, we find that
the Revision Risk already exceeds its pre-pandemic 95th percentile for four
of the eight macroeconomic series, and for all eight series three months
later.
This experiment shows that \SGP{} exhibits significant uncertainty well
before the corresponding X-13 revisions become visible, which for several
series do not peak until two to three years after the shock.
By treating seasonality as a probabilistic concept, the method converts the
reliability of the current decomposition from a retrospective diagnostic
into a quantity that can be monitored in real time, which is especially
useful in high-volatility regimes.

There are limitations to the proposed framework.
One limitation, inherent to all data-driven approaches to seasonal
structure discovery, is non-identifiability of latent seasonal signals,
even in the presence of strong periodicity.
In \SGP, identification is aided by two modeling choices.
First, the prior distribution over Gaussian process covariance kernels, and
second, a normalization step that specifically assigns the average seasonal
level over a fixed window to the nonseasonal component.
A second limitation is that the computational cost of Gaussian process
modeling scales cubically in the number of observations, which imposes
limits on the length of a time series that can be efficiently handled.
Instead of using a full Gaussian process, it is possible to instead
leverage sparse approximations~\citep{Liu2020} with a lower computational
overhead on large datasets, although these methods trade-off
computational cost with modeling accuracy.
A third limitation is the restriction of our framework to handling
a single series at a time.
By extending the method to jointly model multiple time series, it may be
possible to obtain improved decompositions by discovering the
cross-covariance structure between economic signals whose seasonal and
nonseasonal components are highly informative of one another.

\clearpage

\section*{Acknowledgments}
F.~Saad acknowledges support by the National Science Foundation under Award No.~2311983.
Any opinions, findings and conclusions or recommendations expressed in this
material are those of the authors and do not necessarily reflect the views
of the National Science Foundation.

\section*{Declaration of Artificial Intelligence Usage}
The evaluation pipeline used for the studies in
\cref{sec:simulation,sec:empirical}, as well as
\crefrange{fig:paired-levels-spectra-part1}{fig:seasonal-component-part2}
in \cref{appx:frequency}, were developed with programming
assistance from GPT-5.6 Codex.
The underlying AutoGP.jl library is maintained without AI assistance.
A draft of the manuscript was analyzed by GPT-5.6, which surfaced
suggestions for improving the clarity and presentation.
The authors reviewed these items, implemented a subset of those
that they deemed appropriate, and take responsibility for the contents of
this article.

\section*{Data Availability Statement}

A replication package for the analyses in
\cref{sec:simulation,sec:empirical} is available at \doi{10.5281/zenodo.22818922}.

\printbibliography

@book{BoxJenkins:TimeSeries:1976,
author       = {Box, George E. P. and Jenkins, Gwilym M.},
title        = {Time Series Analysis: Forecasting and Control},
publisher    = {Holden-Day},
address      = {San Francisco, CA},
year         = {1976}
}

@book{Brown:TimeSeries:1963,
author       = {Brown, Robert G.},
title        = {Smoothing, Forecasting and Prediction of Discrete Time Series},
publisher    = {Prentice-Hall},
address      = {Englewood Cliffs, NJ},
year         = {1963}
}

@incollection{Granger:1978,
author       = {Clive W. J. Granger},
title        = {Seasonality: Causation, Interpretation, and Implications},
booktitle    = {Seasonal Analysis of Economic Time Series},
editor       = {Arnold Zellner},
publisher    = {U.S. Department of Commerce, Bureau of the Census},
year         = {1978},
pages        = {33--46}
}

@article{GhyselsPerron1990SeasonalUnitRoot,
author       = {Eric Ghysels and Pierre Perron},
title        = {The Effect of Seasonal Adjustment Filters on Tests for a Unit Root},
journal      = {Journal of Econometrics},
volume       = {55},
number       = {1--2},
year         = {1993},
doi          = {10.1016/0304-4076(93)90004-O},
}

@book{harvey1989forecasting,
title        = {Forecasting, Structural Time Series Models and the Kalman Filter},
author       = {Harvey, Andrew C.},
year         = {1989},
publisher    = {Cambridge University Press},
doi          = {10.1017/CBO9781107049994},
}

@article{Micchelli:2006,
author       = {Micchelli, Charles A. and Xu, Yuesheng and Zhang, Haizhang},
title        = {Universal Kernels},
journal      = {Journal of Machine Learning Research},
volume       = {7},
pages        = {2651--2667},
year         = {2006},
url          = {http://jmlr.org/papers/v7/micchelli06a.html}
}

@techreport{X13RefManual:2024,
author       = {{U.S. Census Bureau}},
title        = {{X-13ARIMA-SEATS} Reference Manual},
institution  = {{Center for Statistical Research and Methodology, U.S. Census Bureau}},
type         = {Software Documentation},
number       = {Version 1.1},
address      = {Washington, DC},
month        = apr,
year         = {2025},
url          = {https://www2.census.gov/software/x-13arima-seats/x13as/unix-linux/documentation/docx13as.pdf},
}

@phdthesis{Saad:Dissertation:22,
author       = {Saad, F. A. K.},
title        = {Scalable Structure Learning, Inference, and Analysis with Probabilistic Programs},
school       = {Massachusetts Institute of Technology},
year         = {2022}
}

@inproceedings{SaadEtAl:ICML:23,
title        = {Sequential {Monte} {Carlo} Learning for Time Series Structure Discovery},
author       = {Saad, Feras A. and Patton, Brian J. and Hoffmann, Matthew D. and Saurous, Rif A. and Mansinghka, V. K.},
booktitle    = {Proceedings of the 40th International Conference on Machine Learning},
series       = {Proceedings of Machine Learning Research},
volume       = {202},
pages        = {29473--29489},
year         = {2023},
publisher    = {PMLR}
}

@inproceedings{CusumanoTownerEtAl:PLDI:19,
title        = {{Gen}: A General-Purpose Probabilistic Programming System with Programmable Inference},
author       = {Cusumano-Towner, Marco F. and Saad, Feras A. and Lew, Alexander K. and Mansinghka, Vikash K.},
booktitle    = {Proceedings of the 40th ACM SIGPLAN Conference on Programming Language Design and Implementation},
pages        = {221--236},
year         = {2019},
publisher    = {ACM},
doi          = {10.1145/3314221.3314642}
}

@article{SaadEtAl:POPL:19,
title        = {Bayesian Synthesis of Probabilistic Programs For Automatic Data Modeling},
author       = {Saad, Feras A. and Cusumano-Towner, Marco F. and Schaechtle, Ulrich and Rinard, Martin C. and Mansinghka, Vikash K.},
journal      = {Proceedings of the ACM on Programming Languages},
volume       = {3},
number       = {POPL},
pages        = {37:1--37:32},
eid          = {37},
pagetotal    = {32},
month        = jan,
year         = {2019},
doi          = {10.1145/3290350}
}

@book{ChopinEtAl:2020,
title        = {An Introduction to Sequential Monte Carlo},
author       = {Chopin, Nicolas and Papaspiliopoulos, Omiros},
series       = {Springer Series in Statistics},
publisher    = {Springer},
address      = {Cham},
year         = {2020},
doi          = {10.1007/978-3-030-47845-2}
}

@inproceedings{NeklyudovEtAl:ICML:2020,
title        = {Involutive {MCMC}: A Unifying Framework},
author       = {Neklyudov, Kirill and Welling, Max and Egorov, Evgenii and Vetrov, Dmitry},
booktitle    = {Proceedings of the 37th International Conference on Machine Learning},
series       = {Proceedings of Machine Learning Research},
volume       = {119},
pages        = {7273--7282},
year         = {2020},
publisher    = {PMLR}
}

@book{Rasmussen:2006,
title        = {Gaussian Processes for Machine Learning},
author       = {Rasmussen, Carl E. and Williams, Christopher K. I.},
year         = {2006},
publisher    = {MIT Press},
address      = {Cambridge, MA}
}

@book{Liu:2004,
author       = {Liu, Jun S.},
title        = {Monte {Carlo} Strategies in Scientific Computing},
series       = {Springer Series in Statistics},
publisher    = {Springer-Verlag},
address      = {New York},
year         = {2004},
doi          = {10.1007/978-0-387-76371-2}
}

@article{Watson:1987,
author       = {Watson, Mark W.},
title        = {Uncertainty in Model-Based Seasonal Adjustment Procedures and Construction of Minimax Filters},
journal      = {Journal of the American Statistical Association},
volume       = {82},
number       = {398},
pages        = {395--408},
month        = jun,
year         = {1987},
doi          = {10.1080/01621459.1987.10478442}
}

@article{Wright2013,
author       = {Wright, Jonathan H.},
title        = {Unseasonal Seasonals?},
journal      = {Brookings Papers on Economic Activity},
year         = {2013},
volume       = {44},
number       = {2},
pages        = {65--126},
doi          = {10.1353/eca.2013.0017},
}

@book{Robert:2004,
author       = {Robert, Christian P. and Casella, George},
title        = {Monte {Carlo} Statistical Methods},
edition      = 2,
year         = {2004},
publisher    = {Springer},
address      = {New York},
doi          = {10.1007/978-1-4757-4145-2}
}

@inproceedings{Duvenaud:ICML:2013,
title        = {Structure Discovery in Nonparametric Regression Through Compositional Kernel Search},
author       = {Duvenaud, David and Lloyd, James and Grosse, Roger and Tenenbaum, Joshua B. and Ghahramani, Zoubin},
booktitle    = {Proceedings of the 30th International Conference on Machine Learning},
series       = {Proceedings of Machine Learning Research},
volume       = {28},
pages        = {1166--1174},
year         = {2013},
publisher    = {PMLR},
}

@incollection{MacKay1998,
author       = {MacKay, David J. C.},
title        = {Introduction to Gaussian Processes},
booktitle    = {Neural Networks and Machine Learning},
editor       = {Bishop, Christopher M.},
series       = {NATO ASI Series F: Computer \& Systems Sciences},
volume       = {168},
pages        = {133--166},
publisher    = {Springer},
address      = {Berlin},
year         = {1998}
}

@article{thomson1982,
author       = {Thomson, David J.},
title        = {Spectrum Estimation and Harmonic Analysis},
journal      = {Proceedings of the IEEE},
year         = {1982},
volume       = {70},
number       = {9},
pages        = {1055--1096},
doi          = {10.1109/PROC.1982.12433}
}

@book{percival1993,
author       = {Percival, Donald B. and Walden, Andrew T.},
title        = {Spectral Analysis for Physical Applications: Multitaper and Conventional Univariate Techniques},
publisher    = {Cambridge University Press},
address      = {Cambridge},
year         = {1993},
doi          = {10.1017/CBO9780511622762}
}

@article{babadi2014,
author       = {Babadi, Behtash and Brown, Emery N.},
title        = {A Review of Multitaper Spectral Analysis},
journal      = {IEEE Transactions on Biomedical Engineering},
year         = {2014},
volume       = {61},
number       = {5},
pages        = {1555--1564},
doi          = {10.1109/TBME.2014.2311996}
}

@article{Nerlove:1964,
author       = {Nerlove, Marc},
title        = {Spectral Analysis of Seasonal Adjustment Procedures},
journal      = {Econometrica},
year         = {1964},
volume       = {32},
number       = {3},
pages        = {241--286},
doi          = {10.2307/1913037}
}

@article{RudebuschEtAl:2015,
author       = {Rudebusch, Glenn D. and Wilson, Daniel and Mahedy, Tim},
title        = {The Puzzle of Weak First-Quarter {GDP} Growth},
journal      = {FRBSF Economic Letter},
year         = {2015},
number       = {2015-16},
month        = may,
note         = {Federal Reserve Bank of San Francisco},
url          = {https://www.frbsf.org/economic-research/publications/economic-letter/2015/may/weak-first-quarter-gdp-growth-residual-seasonality/}
}

@techreport{Lunsford:2017,
author       = {Lunsford, Kurt G.},
title        = {Lingering Residual Seasonality in {GDP} Growth},
institution  = {Federal Reserve Bank of Cleveland},
type         = {Economic Commentary},
number       = {2017-06},
year         = {2017},
address      = {Cleveland, OH},
doi          = {10.26509/frbc-ec-201706}
}

@techreport{ConsolvoLunsford:2019,
author       = {Consolvo, Victoria and Lunsford, Kurt G.},
title        = {Residual Seasonality in GDP Growth Remains after Latest BEA Improvements},
institution  = {Federal Reserve Bank of Cleveland},
type         = {Economic Commentary},
year         = {2019},
number       = {2019-05},
address      = {Cleveland, OH},
doi          = {10.26509/frbc-ec-201905}
}

@techreport{Lunsford:2025,
author       = {Lunsford, Kurt G.},
title        = {Residual Seasonality in Five Measures of {PCE} Inflation},
institution  = {Federal Reserve Bank of Cleveland},
type         = {Economic Commentary},
number       = {2025-03},
year         = {2025},
address      = {Cleveland, OH},
doi          = {10.26509/frbc-ec-202503}
}

@techreport{BrysonCornwall:2026,
author       = {Bryson, Carter and Cornwall, Gary},
title        = {The Seasons They Are A-Changin': A Century of Definitions and a Way Forward},
institution  = {U.S.~Bureau of Economic Analysis},
type         = {Working Paper},
number       = {WP2026-14},
address      = {Washington, DC},
year         = {2026},
month        = jun,
doi          = {10.66137/ZHRT5937},

}

@report{Findley:2017,
author       = {Findley, David F. and Lytras, Demetra P. and McElroy, Tucker S.},
title        = {Detecting Seasonality in Seasonally Adjusted Monthly Time Series},
institution  = {Center for Statistical Research \& Methodology, U.S. Census Bureau},
type         = {Research Report Series (Statistics)},
number       = {2017-03},
location     = {Washington, D.C.},
date         = {2017-02-13},
url          = {https://www.census.gov/content/dam/Census/library/working-papers/2017/adrm/rrs2017-03.pdf}
}

@misc{BLSSeasonalAdjustment:2026,
author       = {{U.S. Bureau of Labor Statistics}},
title        = {Seasonal Adjustment Methodology for National Labor Force Statistics},
url          = {https://www.bls.gov/cps/seasonal-adjustment-methodology.htm},
year         = {2026},
note         = {Accessed September 2026}
}

@article{Liu2020,
author       = {Liu, Haitao and Ong, Yew-Soon and Shen, Xiaobo and Cai, Jianfei},
journal      = {IEEE Transactions on Neural Networks and Learning Systems},
title        = {When Gaussian Process Meets Big Data: A Review of Scalable GPs},
year         = {2020},
volume       = {31},
number       = {11},
pages        = {4405--4423},
doi          = {10.1109/TNNLS.2019.2957109},
}

@techreport{MonsellFindley1984,
author       = {Monsell, Brian C. and Findley, David F.},
title        = {Techniques for Determining if a Seasonal Time Series Can Be Seasonally Adjusted Reliably by a Given Seasonal Adjustment Methodology},
institution  = {U.S. Bureau of the Census, Statistical Research Division},
type         = {SRD Research Report},
number       = {CENSUS/SRD/RR-84/14},
address      = {Washington, D.C.},
month        = aug,
year         = {1984},
}

@report{BLS2025EmploymentJuly,
author       = {{U.S. Bureau of Labor Statistics}},
title        = {The Employment Situation---July 2025},
type         = {Economic News Release},
number       = {USDL-25-1202},
institution  = {U.S. Bureau of Labor Statistics},
date         = {2025-08-01},
url          = {https://www.bls.gov/news.release/archives/empsit_08012025.htm}
}

@online{Mutikani2025,
author       = {Mutikani, Lucia},
title        = {{US} Labor Market Cracks Widen as Job Growth Hits Stall Speed},
organization = {Reuters},
date         = {2025-08-01},
url          = {https://www.reuters.com/world/us/us-labor-market-cracks-widen-job-growth-hits-stall-speed-2025-08-01/},
urldate      = {2026-09-17},
}
\pagebreak

\appendix
\section{Gaussian Process Models}
\label{appx:gp}

\subsection{Model Definition}
\label{appx:gp-model}

We briefly overview Gaussian process models for time series data.
A Gaussian process over an index set $\mathbb{T}$ is a family
$\langle X(t) : t \in \mathbb{T} \rangle$ of random variables
such that for any finite tuple
$\mathbf{t} = (t_1 ,\dots,t_n) \in \mathbb{T}^n$
of $n \ge 1$ time indices, the random vector
$\mathbf{X}(\mathbf{t}) \defas \left(X(t_1), \dots, X(t_n)\right)$
follows a multivariate Gaussian distribution:
\begin{gather}
  \mathbf{X}(\mathbf{t}) =
  \begin{pmatrix} X(t_1) \\ \vdots \\ X(t_n) \end{pmatrix}
  \sim \mathcal{N} \left( \mu(\mathbf{t}), K(\mathbf{t},\mathbf{t}) \right),
\end{gather}
where $\mu(\mathbf{t}) \defas (\mu(t_1), \ldots, \mu(t_n))$, with
$\mu(t) \defas \mathbb{E}[X(t)]$, is the mean vector, and $K(\mathbf{t},\mathbf{t})$
is the $n \times n$ covariance matrix with entries $K_{ij} \defas k(t_i, t_j) \defas
\mathbb{E}\left[(X(t_i) - \mu(t_i))(X(t_j) - \mu(t_j))\right]$ for all $1 \le i,j \le n$.
Letting $\mathbf{x}(\mathbf{t}) \defas (x(t_1), \ldots, x(t_n)) \in \mathbb{R}^n$
denote a possible realization of $\mathbf{X}(\mathbf{t})$, the log probability density is given by
\begin{equation}
  \ln p(\mathbf{x}(\mathbf{t}))
    = -\tfrac{1}{2} \left[
      (\mathbf{x}(\mathbf{t}) - \mu(\mathbf{t}))^\intercal
        K(\mathbf{t},\mathbf{t})^{-1}
        (\mathbf{x}(\mathbf{t}) - \mu(\mathbf{t}))
        \right]
    - \tfrac{1}{2} \ln \det\!\left(K(\mathbf{t},\mathbf{t})\right)
    - \tfrac{n}{2} \ln (2\pi).
\end{equation}
Gaussian distributions are closed under conditioning.
In particular,
conditioned on the event $\mathbf{X}(\mathbf{t}) = \mathbf{x}(\mathbf{t})$,
the posterior distribution of
$\mathbf{X}(\mathbf{t}')$ at $n'$ new time points
$\mathbf{t}' = \left(t'_1, \ldots, t'_{n'}\right) \in \mathbb{T}^{n'}$,
is itself multivariate Gaussian:
\begin{align}
\mathbf{X}(\mathbf{t}') \;\big\vert\; \mathbf{X}(\mathbf{t}) = \mathbf{x}(\mathbf{t}) &\sim \mathcal{N}\left(\mu_{\text{post}}(\mathbf{t}'), K_{\text{post}}(\mathbf{t}', \mathbf{t}')\right), \\
\mu_{\text{post}}(\mathbf{t}') &\defas \mu(\mathbf{t}') + K(\mathbf{t}', \mathbf{t}) K(\mathbf{t}, \mathbf{t})^{-1} (\mathbf{x}(\mathbf{t}) - \mu(\mathbf{t})), \label{eq:gp-postx-mu} \\
K_{\text{post}}(\mathbf{t}', \mathbf{t}') &\defas K(\mathbf{t}', \mathbf{t}') - K(\mathbf{t}', \mathbf{t}) K(\mathbf{t}, \mathbf{t})^{-1} K(\mathbf{t}, \mathbf{t}'). \label{eq:gp-postx-cov}
\end{align}

\paragraph{Measurement Noise.}
Observation noise data is modeled by using a stochastic process
$\mathbf{Y}$ that is the
the sum of the latent GP and i.i.d.~Gaussian innovations, i.e.,
$\mathbf{Y}(\mathbf{t}) \defas \mathbf{X}(\mathbf{t}) + \boldsymbol{\epsilon}(\mathbf{t})$,
where $\epsilon(t) \sim \mathcal{N}(0, \eta)$ for $t \in \mathbb{T}$.
As the distribution of $\mathbf{Y}(\mathbf{t})$ is Gaussian with covariance
matrix $K(\mathbf{t}, \mathbf{t}) + \eta I_n$,
the log probability density is given by
\begin{equation}
  \ln p(\mathbf{y}(\mathbf{t}))
    = -\tfrac{1}{2} \left[
      (\mathbf{y}(\mathbf{t}) - \mu(\mathbf{t}))^\intercal
        \left[K(\mathbf{t},\mathbf{t})+ \eta I_n\right]^{-1}
        (\mathbf{y}(\mathbf{t}) - \mu(\mathbf{t}))
        \right]
    - \tfrac{1}{2} \ln \det\!\left(K(\mathbf{t},\mathbf{t})+ \eta I_n\right)
    - \tfrac{n}{2} \ln (2\pi).
  \label{eq:gp-x-pdf}
\end{equation}
%
Conditioning on noisy observations $\mathbf{Y}(\mathbf{t}) = \mathbf{y}(\mathbf{t})$
yields the posterior distribution over the latent process:
\begin{align}
\mathbf{X}(\mathbf{t}') \;\big\vert\; \mathbf{Y}(\mathbf{t}) = \mathbf{y}(\mathbf{t}) &\sim \mathcal{N}\left(\mu_{\text{post}}(\mathbf{t}'), K_{\text{post}}(\mathbf{t}', \mathbf{t}')\right), \\
\mu_{\text{post}}(\mathbf{t}') &\defas \mu(\mathbf{t}') + K(\mathbf{t}', \mathbf{t}) [K(\mathbf{t}, \mathbf{t}) + \eta I_n]^{-1} (\mathbf{y}(\mathbf{t}) - \mu(\mathbf{t})), \label{eq:gp-post-y-mu} \\
K_{\text{post}}(\mathbf{t}', \mathbf{t}') &\defas K(\mathbf{t}', \mathbf{t}') - K(\mathbf{t}', \mathbf{t})\left[K(\mathbf{t}, \mathbf{t}) + \eta I_n\right]^{-1} K(\mathbf{t}, \mathbf{t}'). \label{eq:gp-post-y-cov}
\end{align}

\subsection{Prior Mean}
\label{appx:gp-mean}

It is standard in Gaussian process modeling to assume the prior mean
satisfies $\mu \equiv 0$ \citep[Chapter 2]{Rasmussen:2006}.
This convention is largely without loss of generality for our purposes,
because systematic structure can be represented using the covariance
kernel rather than the mean function.
For example, suppose a Gaussian process $X$ with covariance kernel
$k$ has an affine mean function $\mu(t) = A(t-c) + B$,
with unknown slope $A$ and intercept $B$:
\begin{align}
A &\sim \mathcal{N}(0, \sigma_A^2) \\
B &\sim \mathcal{N}(0, \sigma_B^2) \\
X(\cdot) \mid A, B &\sim \mathrm{GP}(A(t-c) + B, k).
\end{align}
Marginally, $X$ is equivalent to a mean-zero Gaussian process
with covariance kernel $k + k'$:
\begin{align}
X(\cdot) \sim \mathrm{GP}\left(\mu(t) =0,\, k + k' \right)
	&& k'(t,t') \defas \sigma_A^2(t-c)(t'-c)+\sigma_B^2.
\end{align}
Here, the unknown parameters $A$ and $B$ are analytically marginalized out,
while the hyperparameters $\sigma_A$, $\sigma_B$, and $c$
become parameters of the covariance kernel $k'$.
In general, affine and other structures with unknown parameters can often
be represented through a composite covariance kernel, even when the prior
mean is set to zero.
On the other hand, deterministic components with known parameters, such as
fixed offsets or calendar corrections, can either be encoded in the mean
function or removed by preprocessing as in \cref{appx:preproc}.

As deterministic components are typically not known a priori in our model
discovery setting, we set the prior mean to zero and let level, trend,
periodicity, and other temporal structures be inferred using the kernel
grammar and posterior inference.
It should be noted that although the prior mean is zero, posterior
summaries of the Gaussian process such as means,
medians, and quantiles, become nonzero after conditioning
on data.

\subsection{Covariance Kernels}
\label{appx:gp-kernels}

A large number of Gaussian process covariance kernels are described in
\citet{MacKay1998} and \citet{Rasmussen:2006}.
Our model discovery framework makes use of the following subset.

\paragraph{\gpC.}
\begin{equation}
k_\pC(t, t') \defas \theta_\texttt{C},
\end{equation}
where
$\theta_\texttt{C} > 0$ is the constant variance value. This kernel
produces a full covariance matrix with every entry equal to
$\theta_\texttt{C}$, implying that all function values are perfectly
correlated. Realizations from a Gaussian process with this kernel are
horizontal lines.

\paragraph{\gpLin.}
\begin{equation}
k_\pLin(t, t') = \theta_{\pLin,2} (t - \theta_{\pLin,1})(t' - \theta_{\pLin,1}),
\end{equation}
where $\theta_{\pLin,1}$ is the intercept that centers the input
domain, and $\theta_{\pLin,2}$ controls the slope or scaling of the
linear trend. This kernel generates covariance matrices whose entries
grow with distance from the intercept, and draws from a Gaussian process
with this kernel are sloped lines.

\paragraph{\gpGe.}
\begin{equation}
k_{\pGe}(t, t')
  \defas \theta_{\pGe,3} \exp\left(-\left( \frac{\abs{t-t'}}{\theta_{\pGe,1}}  \right)^{\theta_{\pGe,2}} \right)
\end{equation}
where $\theta_{\pGe,1}$ is the lengthscale,
$\theta_{\pGe,2} \in (0,2]$ is the shape parameter (often denoted $\gamma$),
and $\theta_{\pGe,3}$ is the amplitude.
It allows for modeling functions that vary in smoothness
across different time scales.
Points that are close together are strongly
correlated while points farther apart become less correlated. The rate at
which this correlation decays depends on a parameter $\theta_{\pGe,2}$:
larger values produce smoother function draws, while smaller values allow more
rapidly varying functions.
Setting $\theta_{\pGe,2} =2$ gives an equivalent parameterization
of the \gpSe{} kernel.

\paragraph{\gpPer.}
\begin{equation}
  k_\pPer(t, t')
  \defas \theta_{\pPer,3}
  \exp\left( - {2 \sin^2\left( \displaystyle\frac{\pi}{\theta_{\pPer,2}} \left|t - t'\right| \right)}\Big/{\theta_{\pPer,1}^2} \right)
\end{equation}
where $\theta_{\pPer,1}$ is the lengthscale controlling smoothness
within each period, $\theta_{\pPer,2}$ is the period of repetition,
and $\theta_{\pPer,3}$ is the amplitude.
The kernel can be understood by warping the time points $t$
to the unit circle using
$\theta(t) \defas (\cos(2\pi/\theta_{\pPer,2}t), \sin(2\pi/\theta_{\pPer,2}t))$
and then applying the \gpSe{} kernel to inputs $\theta(t), \theta(t')$.
Further description is given in \cref{sec:seasonality-kernel}.

\subsection{Proof of \titlezcref{Prop:PostComponDecomp}}
\label{appx:gp-proof}

\Crefrange{eq:train-vars}{eq:train-cross}
are derived by conditioning on the data $\mathbf{Y}$ using the
formula
\begin{equation}
\Var{\mathbf{Z}_1 \mid \mathbf{Z}_2} = \boldsymbol{\Sigma}_{11} - \boldsymbol{\Sigma}_{12} \boldsymbol{\Sigma}_{22}^{-1} \boldsymbol{\Sigma}_{21}
\end{equation}
where $\mathbf{Z}_1 = [\mathbf{S}, \mathbf{U}]$
and $\mathbf{Z}_2 = \mathbf{Y}$:
\begin{align}
  \Var{\begin{bmatrix} \mathbf{S} \\ \mathbf{U} \end{bmatrix} \;\middle|\; \mathbf{Y}}
  &= \underbrace{\begin{bmatrix} K_m^{\mathrm{per}} & \mathbf{0} \\ \mathbf{0} & K_m^{\mathrm{non}} \end{bmatrix}}_{\boldsymbol{\Sigma}_{11}}
    - \underbrace{\begin{bmatrix} K_m^{\mathrm{per}} \\ K_m^{\mathrm{non}} \end{bmatrix}}_{\boldsymbol{\Sigma}_{12}}
      \underbrace{(K+\eta I)^{-1}}_{\boldsymbol{\Sigma}_{22}^{-1}}
      \underbrace{\begin{bmatrix} K_m^{\mathrm{per}} & K_m^{\mathrm{non}} \end{bmatrix}}_{\boldsymbol{\Sigma}_{21}}
  \\
  &= \begin{bmatrix} K_m^{\mathrm{per}} & \mathbf{0} \\ \mathbf{0} & K_m^{\mathrm{non}}
  \end{bmatrix} -
    \begin{bmatrix}
      K_m^{\mathrm{per}}(K+\eta I)^{-1}K_m^{\mathrm{per}} & K_m^{\mathrm{per}}(K+\eta I)^{-1}K_m^{\mathrm{non}} \\
      K_m^{\mathrm{non}}(K+\eta I)^{-1}K_m^{\mathrm{per}} & K_m^{\mathrm{non}}(K+\eta I)^{-1}K_m^{\mathrm{non}}
    \end{bmatrix}
  \\
  &= \begin{bmatrix}
    K_m^{\mathrm{per}} - K_m^{\mathrm{per}}(K+\eta I)^{-1}K_m^{\mathrm{per}} & -K_m^{\mathrm{per}}(K+\eta I)^{-1}K_m^{\mathrm{non}} \\
    -K_m^{\mathrm{non}}(K+\eta I)^{-1}K_m^{\mathrm{per}} & K_m^{\mathrm{non}} - K_m^{\mathrm{non}}(K+\eta I)^{-1}K_m^{\mathrm{non}}
  \end{bmatrix}
  \\
  &= \begin{bmatrix}
    \Var{\mathbf{S}\mid\mathbf{Y}} & \mathrm{Cov}(\mathbf{S},\mathbf{U}\mid\mathbf{Y}) \\
    \mathrm{Cov}(\mathbf{U},\mathbf{S}\mid\mathbf{Y}) & \Var{\mathbf{U}\mid\mathbf{Y}}
  \end{bmatrix}.
\end{align}
The conclusion follows.


\section{Data Description}
\label{appx:data}

This appendix describes the data used for the empirical
evaluation in \cref{sec:empirical}, which are summarized in \cref{table:series}.

The empirical analysis uses eight monthly U.S. macroeconomic indicators
spanning labor markets, consumer prices, industrial production, retail
spending, residential construction, manufacturing demand, and business
inventories.
For each indicator, the not seasonally adjusted (NSA) series is the object
modeled by AutoGP, while the corresponding official seasonally adjusted (SA)
series provides a descriptive benchmark.
Data were retrieved from Fred on July~21, 2026.
The sample ends in September~2025, the latest uninterrupted month
common to all 16 NSA and SA series.
The sample begins in January~1990 where available.
Retail Sales and Business Inventories begin in January~1992, and
Durable-Goods Orders begins in February~1992.

Payroll Employment, \texttt{PAYNSA}, measures employees on total nonfarm
payrolls in thousands of persons.
The two price indexes provide deliberately different seasonal environments:
\texttt{CPILFENS} is the broad CPI excluding food and energy, whereas
\texttt{CUUR0000SAA2} is the narrower CPI for women's and girls' apparel, which
has a much stronger recurring seasonal pattern.
\texttt{IPB50001N} measures total industrial production, and
\texttt{RSAFSNA} measures nominal advance retail and food-services sales.
\texttt{HOUSTNSA} measures the monthly number of privately owned housing units
started, \texttt{UMDMNO} measures manufacturers' new orders for durable goods,
and \texttt{TOTBUSIMNSA} measures inventories held by manufacturers,
wholesalers, and retailers.
One scaling difference is important when comparing levels.
The official Housing Starts series, \texttt{HOUST}, is reported at a seasonally
adjusted annual rate, while \texttt{HOUSTNSA} is a monthly count in thousands
of units.
Consequently, the official SA level is approximately 12 times the corresponding
seasonally adjusted monthly count.
This constant scaling difference cancels from monthly log growth and therefore
does not affect the frequency-domain or NSA-minus-SA growth comparisons below.

\section{Data Preprocessing}
\label{appx:preproc}

Monthly economic series can be affected by variation in the number and
composition of weekdays within each month.
Flow variables are especially sensitive to the number of weekdays in the month.
In the forecasting exercise from \cref{sec:empirical-forecast},
we remove deterministic trading-day effects from the calendar-sensitive
series before estimating AutoGP for four series where calendar composition
has been demonstrated as relevant: Retail Sales, Industrial Production,
Housing Starts, and Durable-Goods Orders.
For each month $t$ we count the occurrences of each weekday, letting $D_j(t)$
denote the number of weekday $j$.
Taking Sunday as the omitted baseline, the
trading-day regressors are $T_j(t) \defas{} D_j(t) - D_{\mathrm{Sun}}(t)$,
for $j \in \{\mathrm{Mon},\mathrm{Tue},\mathrm{Wed},\mathrm{Thu},\mathrm{Fri},\mathrm{Sat}\}$.
Let $y_i(t)$ denote the log NSA values of series $i$ at month $t$.
We estimate the log-additive trading-day effect in first differences
\begin{equation}
  \Delta y_i(t)
  \defas \alpha_i
  + \sum_{j\in\set{\mathrm{Mon},\ldots,\mathrm{Sat}}}\mspace{-30mu} \beta_{i,j}\,\Delta T_j(t)
  + \sum_{l=2}^{12} \gamma_{i,l} M_l(t)
  + u_i(t),
  \label{eq:panel-td-reg}
\end{equation}
where the $M_l(t)$ are month dummies.
Differencing removes the trend and the month dummies absorb the seasonal
mean, so the $\hat\beta_{i,j}$ identify the trading-day effect net of both.
We then reconstruct the trading-day component in log levels and subtract it,
\begin{equation}
  \widehat{\mathrm{TD}}_i(t)
  \defas \sum_{j\in\set{\mathrm{Mon},\ldots,\mathrm{Sat}}}\mspace{-30mu} \hat\beta_{i,j}\,T_j(t),
  \qquad
  z_i(t) \defas y_i(t) - \widehat{\mathrm{TD}}_i(t),
  \label{eq:panel-td-component}
\end{equation}
yielding the trading-day-scrubbed log series $z_i(t)$ on which AutoGP is
estimated.

As the Core CPI and Business Inventories series are highly persistent and
trend-dominated, we estimate AutoGP using monthly log differences.
No preprocessing is performed for the two remaining series (Payroll
Employment and Apparel CPI).
The resulting data that AutoGP is estimated on is
\begin{align}
z_i(t) \defas \begin{cases}
  y_i(t) - \widehat{\mathrm{TD}}_i(t) & i = \set*{\begin{aligned}&\mbox{Retail Sales}, \mbox{Industrial Production},\\&\mbox{Housing Starts}, \mbox{Durable-Goods Orders}\end{aligned}},
  \\
  y_i(t) - y_i(t-1) & i \in \set*{\mbox{Core CPI}, \mbox{Business Inventories}},
  \\
  y_i(t) & i \in \set*{\mbox{Payroll Employment}, \mbox{Apparel CPI}}.
  \end{cases}
\end{align}

All forecast evaluation is performed on log-level target $y_i(t)$.
For each series $i$, forecast origin $v$, and forecast horizon $h = 1,\dots,12$,
AutoGP produces predictions $\hat{z}_i(v+h)$.
The predicted log-level series $\hat{y}_i$ is reconstructed as
\begin{align}
\hat{y}_i(v+h) \defas \begin{cases}
  \hat{z}_i(v+h) + \widehat{\mathrm{TD}}_i(v+h) & i \in \set*{\begin{aligned}&\mbox{Retail Sales}, \mbox{Industrial Production},\\&\mbox{Housing Starts}, \mbox{Durable-Goods Orders}\end{aligned}},
  \\
  y_i(v) + \sum_{j=1}^h \hat{z}_i(v+j) & i \in \set*{\mbox{Core CPI}, \mbox{Business Inventories}},
  \\
  \hat{z}_i(v+h) & i \in \set*{\mbox{Payroll Employment}, \mbox{Apparel CPI}}.
  \end{cases}
\end{align}

\section{Frequency and Time--Domain Summaries}
\label{appx:frequency}

This appendix describes the frequency analysis in \cref{sec:empirical-freq}.
For series $i$, let $y^{\mathrm{NSA}}_i(t)$ and $y^{\mathrm{SA}}_i(t)$
denote the published NSA series and its SA counterpart in levels.
The frequency-domain summaries use monthly log growth,
\begin{align}
g^{s}_i(t)
\defas
100\left(\log y^{s}_i(t)-\log y^{s}_i(t-1)\right)
&&
(s \in \set{\mathrm{NSA},\mathrm{SA}}),
\label{eq:data_appendix_log_growth}
\end{align}
which limits domination by low-frequency trend variation and sharpens
concentration at seasonal frequencies.

Spectra are estimated with the DPSS multitaper method of \citet{thomson1982},
which reduces the variance and leakage of a single-taper periodogram
\citep{percival1993,babadi2014}.
For a demeaned log-growth series $g^\circ(t) \defas g(t) - \overline{g}$ of length $N$ and DPSS
taper $v^{(k)}(t)$, the plotted spectrum is the equally weighted average of the
$K$ tapered spectra,
\begin{equation}
\widehat S_{\mathrm{MT}}(f) \defas
  \frac{1}{K} \sum_{k=0}^{K-1}
    \left| \sum_{t=0}^{N-1}v_t^{(k)}g^\circ(t)\,e^{-\mathrm{i}2\pi f t} \right|^2 .
\label{eq:data_appendix_multitaper_spectrum}
\end{equation}
We set $NW=2$ and $K=2NW-1=3$, which resolves the seasonal frequencies
$f_j=j/12$ ($j=1,\ldots,6$) while averaging three approximately uncorrelated
estimates.
The series is demeaned but not prewhitened, and zero-padded before the FFT for
a finer plotting grid.
This transformation does not increase statistical resolution.

\begin{figure}[!htbp]
\centering
\includegraphics[width=\textwidth]{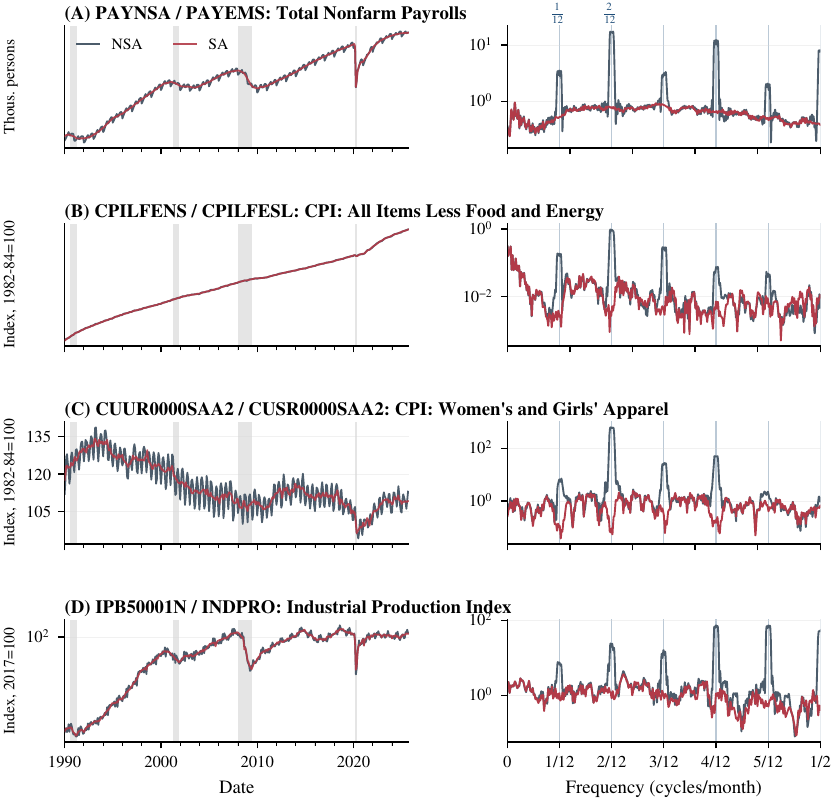}
\caption[Levels and spectra of monthly U.S.~macroeconomic series, part 1]{
Levels and spectra of monthly U.S.~macroeconomic series.
Each row compares the NSA series (dark) with its official SA counterpart (red).
The left column reports levels, using a logarithmic vertical scale where
indicated; shaded regions denote NBER recessions.
The right column reports the DPSS multitaper spectrum of monthly log growth,
$g_t=100\,\Delta\log y_t$, with vertical lines at $j/12$ for
$j=1,\ldots,6$.
\par\smallskip
\emph{Notes.}
Spectra use $NW=2$ and $K=3$, with demeaning, no prewhitening, and a
zero-padded FFT.
The sample is monthly from 1990:01 through 2025:09.
Continued in \cref{fig:paired-levels-spectra-part2}.
}
\label{fig:paired-levels-spectra-part1}
\end{figure}

\begin{figure}[!htbp]
\centering
\includegraphics[width=\textwidth]{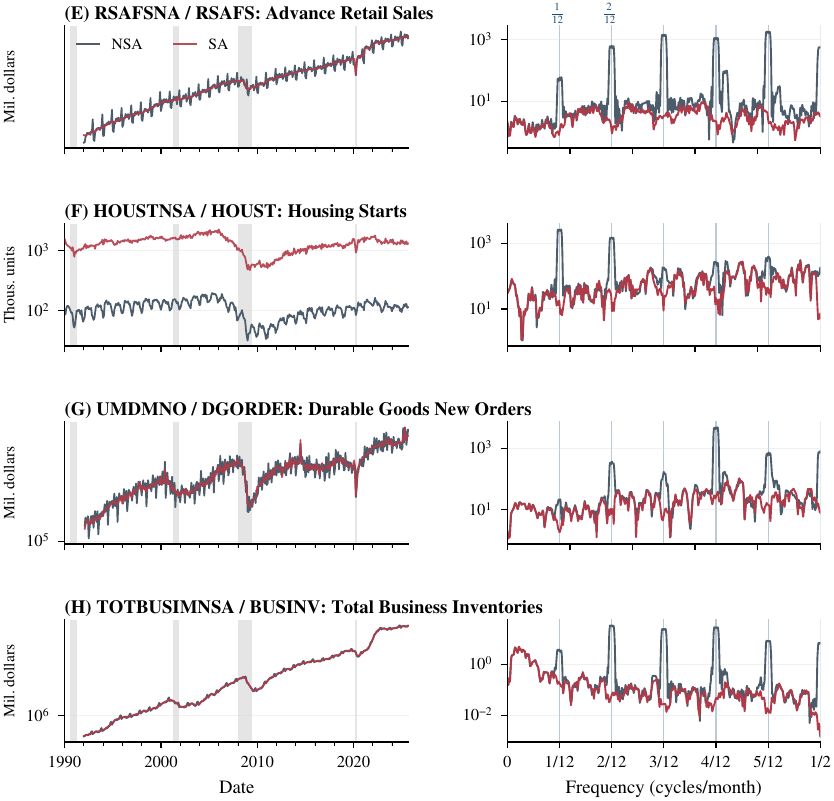}
\caption[Levels and spectra of monthly U.S. macroeconomic series, part 2]{%
Levels and spectra of monthly U.S. macroeconomic series.
The format follows \cref{fig:paired-levels-spectra-part1}.
\par\smallskip
\emph{Notes.}
Housing Starts begins in 1990:01, Retail Sales and Business Inventories begin
in 1992:01, and Durable-Goods Orders begins in 1992:02; all samples end in
2025:09.
The official \texttt{HOUST} series is reported at a seasonally adjusted annual
rate, so the level separation in panel~(F) reflects annualization rather than
residual seasonality; this constant factor cancels from the log-growth
spectrum.
Continued from \cref{fig:paired-levels-spectra-part1}.}
\label{fig:paired-levels-spectra-part2}
\end{figure}

\Cref{fig:paired-levels-spectra-part1,fig:paired-levels-spectra-part2} show
that the series exhibit heterogeneity in the strength and attenuation
of seasonality.
Payroll Employment, Apparel CPI, Industrial Production, Retail Sales,
Durable-Goods Orders, and Business Inventories show pronounced NSA peaks at one
or more seasonal frequencies, with smaller peaks in their SA counterparts.
Core CPI has much less seasonal-frequency power than Apparel CPI.
Housing Starts has exceptionally strong NSA seasonality, and its SA
counterpart retains more seasonal-frequency power than most other SA
series.

\begin{figure}[!htbp]
\centering
\includegraphics[width=\textwidth]{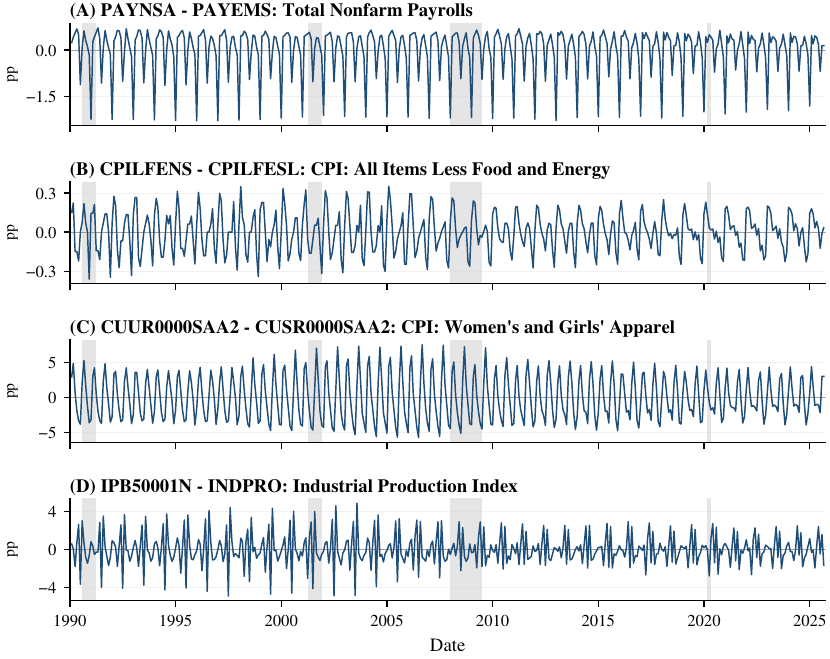}
\caption[Implied official adjustment, part 1]{%
Implied official adjustment: NSA minus SA monthly log growth.
Each panel reports $a_i^{\mathrm{OFF}}(t)$ in percentage points at a monthly
rate.
Shaded regions denote NBER recessions, and the horizontal line denotes no
NSA--SA growth difference.
\par\smallskip
\emph{Notes.}
The sample is monthly from 1990:02 through 2025:09.
The first month is lost when log growth is computed.
Larger absolute values indicate a larger difference between the published NSA
and SA monthly growth rates.
Continued in \cref{fig:seasonal-component-part2}.}
\label{fig:seasonal-component-part1}
\end{figure}

\begin{figure}[!htbp]
\centering
\includegraphics[width=\textwidth]{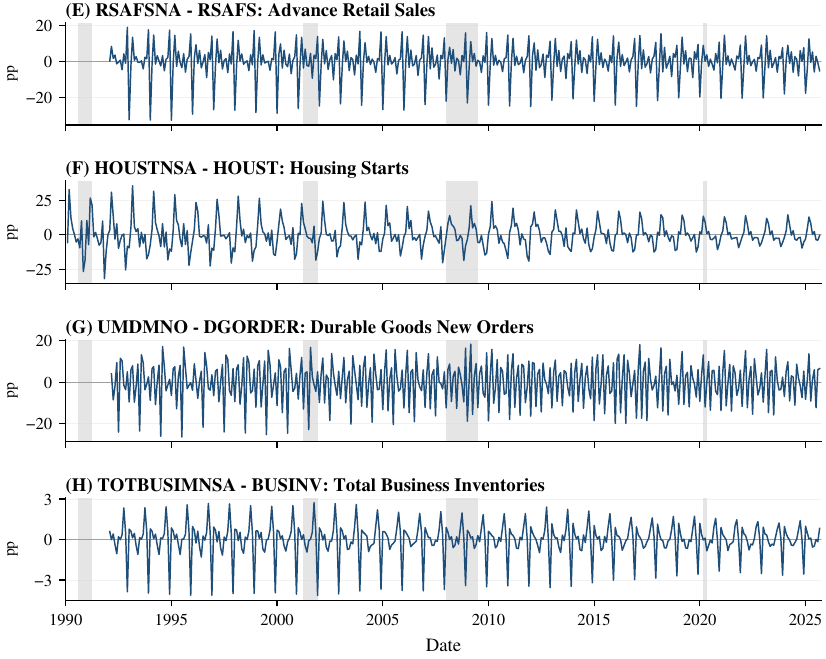}
\caption[Implied official adjustment, part 2]{%
Implied official adjustment: NSA minus SA monthly log growth.
The format follows \cref{fig:seasonal-component-part1}.
\par\smallskip
\emph{Notes.}
The first plotted observations are 1992:02 for Retail Sales and Business
Inventories, 1990:02 for Housing Starts, and 1992:03 for Durable-Goods Orders;
all samples end in 2025:09.
Continued from \cref{fig:seasonal-component-part1}.}
\label{fig:seasonal-component-part2}
\end{figure}

For each indicator,
\cref{fig:seasonal-component-part1,fig:seasonal-component-part2}
show the difference between NSA and SA monthly log
growth,
\begin{equation}
a_i^{\mathrm{OFF}}(t) \defas g^{\mathrm{NSA}}_i(t)-g^{\mathrm{SA}}_i(t),
\label{eq:data_appendix_official_adjustment}
\end{equation}
i.e., the implied official seasonal adjustment.
We avoid calling it a directly observed seasonal component because published
NSA and SA series can also differ through aggregation and other agency
modeling choices.
The plots show that the periodic patterns vary widely across the dataset adjustments
in terms of the locations, magnitudes, and regularity of the seasonal peaks.

\end{document}